\documentclass[11pt]{article}
\usepackage[T1]{fontenc}
\usepackage{lmodern}
\usepackage{amsmath,amssymb,amsthm,mathrsfs}
\usepackage{graphicx,booktabs,float,subcaption,tabularx}
\usepackage[authoryear,round]{natbib}
\usepackage{xcolor}
\usepackage[colorlinks=true,linkcolor=black,citecolor=blue,urlcolor=blue]{hyperref}
\usepackage[margin=1in]{geometry}
\newtheorem{theorem}{Theorem}[section]
\newtheorem{lemma}[theorem]{Lemma}
\newtheorem{proposition}[theorem]{Proposition}

\theoremstyle{remark}
\newtheorem{remark}[theorem]{Remark}
\newcommand{\E}{\mathbb E}
\newcommand{\Q}{\mathbb Q}
\newcommand{\F}{\mathcal F}
\newcommand{\HH}{\mathcal H}
\newcommand{\1}{\mathbf 1}
\newcommand{\norm}[1]{\left\lVert#1\right\rVert}
\newcommand{\pathfigure}[3]{%
 \begin{figure}[htbp]\centering
 \IfFileExists{#1}{\includegraphics[width=#2\linewidth]{#1}}{\fbox{\parbox[c][3cm][c]{.82\linewidth}{\centering Figure file not supplied: \texttt{\detokenize{#1}}}}}
 \caption{#3}\end{figure}}
\title{Valuation Model of Chinese Convertible Bonds Based on Monte Carlo Simulation:\\A Regression-Based Approximation and Error Analysis}
\author{Yu Liu\\School of Science and Engineering, The Chinese University of Hong Kong, Shenzhen\\\texttt{yu130dfb4@gmail.com}}
\date{Revised preprint, November 2026}
\begin{document}
\maketitle
\begin{abstract}
We study a regression-based Monte Carlo method for valuing Chinese convertible bonds with path-dependent call and put triggers and a specified, exogenous issuer reset policy. The holder's exercise decision is expressed as a finite-horizon dynamic-programming recursion. Conditional continuation values are approximated by least-squares regression using the underlying stock price and rolling-window trigger indicators. We compare a global regression specification with regressions fitted separately in four contract-defined price regions. For fixed regression spaces and a well-posed design, we establish convergence of the simulated regression recursion to its finite-dimensional population counterpart and bound the population approximation error by accumulated one-step projection errors. A nested-space argument establishes a smaller or equal \emph{single-step population projection error} for segmented regression on a common target; it does not guarantee a smaller finite-sample pricing error. Historical empirical comparisons on ten AAA-rated Chinese convertible bonds report a mean bond-level relative RMSE of 3.42\% without the put-window features, 3.14\% with them, and 2.96\% under four-region regression. These empirical figures are retained from the original study and require recomputation if the model or implementation changes. A separate nine-period CRR benchmark with a matching exact dynamic-programming reference, 60 replications, and path counts up to 50,000 finds smaller numerical RMSE for the four-region method in this test, while both fixed regression spaces retain nonzero error relative to the exact model value.
\end{abstract}

\section{Introduction}
A convertible bond combines fixed-income and equity-linked payoffs. Chinese convertible bonds (CCBs) commonly include conversion rights, issuer calls, holder puts and conversion-price reset provisions. Their triggers may depend on the number of qualifying stock-price observations within a rolling window. Consequently, a contract's value generally depends on more than the current stock price.

Convertible bond valuation methods include component-based decompositions and integrated numerical valuation. Numerical approaches include simulation \citep{Ammann2008}, lattice models \citep{Milanov2012}, and finite-difference methods \citep{Brennan1978}. The exercise decisions can be treated as a finite-horizon stopping problem once the issuer's policy is specified. For two-sided discretionary exercise rights, a game-option formulation may instead be needed \citep{CrepeyRahal2011,BeveridgeJoshi2011}; this paper deliberately does not claim to solve that more general game.

Regression-based Monte Carlo methods approximate continuation values by projecting discounted future payoffs onto a finite-dimensional function space \citep{LongstaffSchwartz2001,Carriere1996}. Their convergence and error decomposition have been extensively studied \citep{Clement2002,Stentoft2004}. We specialize a regression-based dynamic-programming construction to CCB trigger features and contrast global and four-region regression spaces. The purpose of the theoretical section is to clarify what can be established for this construction under explicit assumptions, not to claim a new general convergence theorem.

The paper proceeds with a contract model and a precise numerical recursion in Section~\ref{sec:model}, theoretical results in Section~\ref{sec:theory}, the originally reported empirical comparisons in Section~\ref{sec:empirical}, an independently reproducible discrete benchmark in Section~\ref{sec:benchmark}, and a trading application in Section~\ref{sec:trading}. The modelling assumptions and limits of the reported evidence are discussed at the end.

\section{Contract model and numerical algorithm}\label{sec:model}
\subsection{Underlying stock and information}
Let $(\Omega,\F,(\F_t)_{t=0}^T,\Q)$ be a filtered probability space on a finite grid of trading dates $t=0,\ldots,T$ with a constant step $\Delta t$ measured in years. Under a specified risk-neutral measure $\Q$, we adopt the geometric Brownian motion model
\begin{equation}\label{eq:sde}
 dS_u=(r-q)S_u\,du+\sigma S_u\,dW_u,
\end{equation}
with constant continuously compounded annual risk-free rate $r$, dividend yield $q$ and annual volatility $\sigma$. Its exact time-grid simulation is
\begin{equation}\label{eq:gbm}
 S_{t+1}^{(i)}=S_t^{(i)}\exp\!\left\{\left(r-q-\tfrac12\sigma^2\right)\Delta t+\sigma\sqrt{\Delta t}\,Z_{t+1}^{(i)}\right\},\qquad Z_{t+1}^{(i)}\sim N(0,1),
\end{equation}
with independent standard normal innovations across paths and dates. The one-step discount factor is $\beta=\exp(-r\Delta t)$. Historical initial information is held fixed when simulating the continuation of one bond from a given valuation date.

For a bond with face value $FV$, write $C_t$ for its \emph{effective} conversion price and $m_t=FV/C_t$ for the conversion ratio. The contractual call and put trigger levels are denoted $k_t$ and $p_t^{\mathrm{tr}}$ (the put \emph{payment} is denoted $P_t$). If the call tests $a_c$ qualifying days in a window of length $n_c$, define
\begin{equation}\label{eq:triggers}
 F_t=\frac{1}{n_c}\sum_{j=0}^{n_c-1}\1_{\{S_{t-j}>k_{t-j}\}},\quad
 Y_t=\frac{1}{n_p}\sum_{j=0}^{n_p-1}\1_{\{S_{t-j}<p_{t-j}^{\mathrm{tr}}\}},\quad
 p_F=\frac{a_c}{n_c},\quad p_Y=\frac{a_p}{n_p}.
\end{equation}
These definitions are a convenient representation of the $a$-out-of-$n$ eligibility rules; eligibility periods and any contractual exclusions must additionally be imposed. The history prior to the valuation date initializes these windows. The \emph{ordered} history, not merely the two proportions $F_t,Y_t$, is retained when generating the next trigger values. Consequently, the stochastic state and filtration include all past observations needed for the contractual rules. Reset-adjusted trigger levels must use the levels in force on the days being tested.

\subsection{Specified issuer policy and optimal stopping}
The holder can elect conversion during a specified conversion period and can exercise a put only when contractually eligible. We treat issuer decisions as a \emph{specified exogenous policy}, rather than optimize issuer exercise adversarially. In particular, an eligible forced call has priority and terminates the bond at its contractual settlement. On a non-call date, the issuer can choose a conversion-price reset with an assumed probability $p\in[0,1]$ when a designated put-related trigger holds. For illustration, the original empirical specification used $p=0.8$; this is a modelling input, not an estimated structural parameter.

Under a reset, a designated floor, for example the greater of the preceding twenty-day stock-price average and the relevant preceding close, is applied only if it is below the existing $C_t$ and if the contract permits a reduction. The resulting conversion ratio affects subsequent choices. The reset and the issuer's immediate call decision are included in the simulated issuer history, which is assumed adapted and independent across simulated full paths. The holder observes any issuer action taking effect before their decision at date $t$. Precise timing matters and must match the actual implementation.

Let $G_t$ denote the largest \emph{eligible} immediate payment to the holder on a non-call date, including $m_tS_t$ when conversion is available and $P_t$ when the put is exercisable. If no voluntary exit is available write $G_t=-\infty$ by convention; in that case the decision operator below is the identity. Define the adapted decision operator
\begin{equation}\label{eq:decision}
 D_t(h)=\begin{cases}
 K_t^{\mathrm{settle}},&\text{if a forced call takes effect at }t,\\
 \max\{G_t,h\},&\text{if at least one holder exit is eligible},\\
 h,&\text{otherwise},
 \end{cases}
\end{equation}
where $K_t^{\mathrm{settle}}$ is the applicable contract settlement (including conversion value when contractually allowed). The forced-call settlement is not a voluntary choice by the holder. Contractual cash coupons, credit risk and transaction frictions are excluded from this \emph{simplified} valuation recursion; they can only be incorporated by changing the state and cash-flow recursion accordingly.

Let $B$ be the terminal redemption payment. The exact value \emph{under the specified issuer policy and model} is the adapted recursion
\begin{equation}\label{eq:bellman}
 V_T=\max\{B,m_TS_T\},\qquad H_t=\E_{\Q}[\beta V_{t+1}\mid\F_t],\qquad V_t=D_t(H_t),\quad t=T-1,\ldots,0.
\end{equation}
This is the holder's finite-horizon optimal stopping recursion with exogenous issuer actions; it is not an unconditional issuer--holder Dynkin-game value. All conditional expectations take account of the full information $\F_t$, including price history and issuer state.

\pathfigure{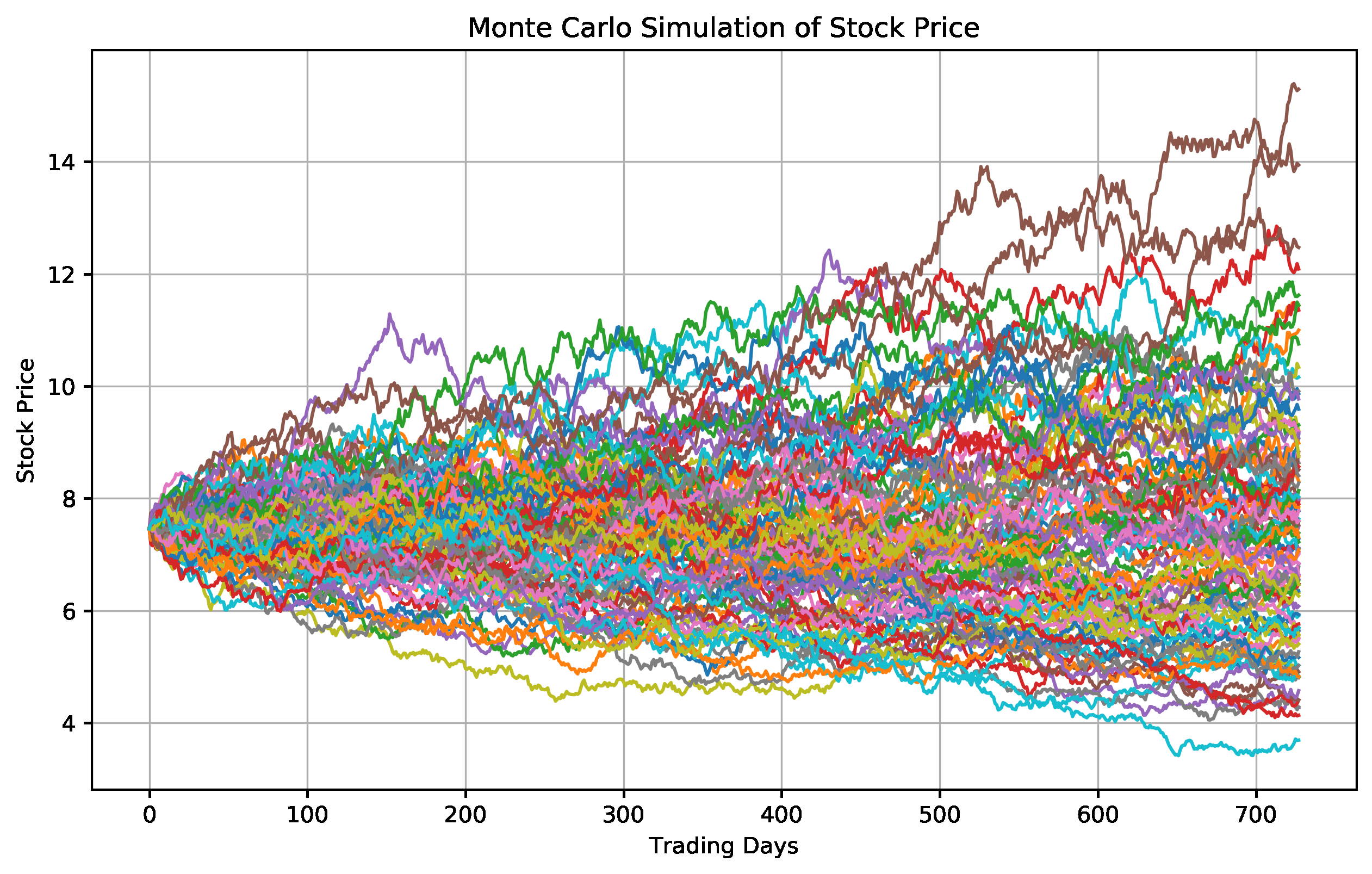}{0.76}{Simulated underlying stock paths for Daqin convertible bond (recovered from the original manuscript).}
\subsection{Regression features and backward numerical recursion}\label{sec:algorithm}
Generate $M$ independent full market-and-issuer histories $i=1,\ldots,M$ from the given initial state. At each intermediate date define the original nine-component feature vector
\begin{equation}\label{eq:features}
 e_t=(S_t,S_t^2,F_t,F_t^2,Y_t,Y_t^2,S_tF_t,S_tY_t,F_tY_t)^{\top}.
\end{equation}
An intercept may be added in new implementations, but the retained historical results use the stated specifications. The labels $F_t,Y_t$ denote regression \emph{features}, not a complete Markov state. Features may be linearly redundant on particular regions; redundant columns are discarded or a Moore--Penrose pseudoinverse is used in such cases.

Initialize $\widehat V_T^{(i)}=V_T^{(i)}$. For each date $t=T-1,\ldots,1$, form responses $y_t^{(i)}=\beta\widehat V_{t+1}^{(i)}$, let $E_t$ have row $(e_t^{(i)})^{\top}$, and solve
\begin{equation}\label{eq:ols}
 \widehat\theta_t\in\arg\min_{b\in\mathbb R^{K_t}}\sum_{i=1}^M\big(y_t^{(i)}-(e_t^{(i)})^{\top}b\big)^2,\qquad
 \widehat V_t^{(i)}=D_t^{(i)}\big((e_t^{(i)})^{\top}\widehat\theta_t\big).
\end{equation}
Here $K_t$ denotes the retained feature dimension and \emph{not} the forced-call settlement. A pseudoinverse selects a solution when the empirical design is rank deficient. At a fixed, known initial state, set
\begin{equation}\label{eq:price}
 \widehat V_0^{(M)}=D_0\!\left(\frac{1}{M}\sum_{i=1}^M\beta\widehat V_1^{(i)}\right).
\end{equation}
At $t=0$ the decision operator is taken to be fixed by the initial contractual information. In particular, the formula reduces to the sample average of discounted path values when no immediate exercise or forced call is available at inception.

\begin{remark}[Which regression algorithm is analyzed?]
Equation~\eqref{eq:ols} propagates the \emph{fitted value} $\widehat V_t^{(i)}$ to the preceding date. Standard policy-based Longstaff--Schwartz implementations may instead propagate the realized future cash flow conditional on the exercise decision. Those versions are related but different numerical algorithms. The results in Section~\ref{sec:theory} apply to the explicit recursion \eqref{eq:ols}--\eqref{eq:price}; the code used for any empirical claim must be checked against it.
\end{remark}

\subsection{Global versus contract-driven segmented regression}
The global method uses the span of the features in \eqref{eq:features}. The segmented method estimates a separate coefficient vector in four stock-price regions,
\begin{equation}\label{eq:regions}
 R_{1,t}=\{S_t>k_t\},\quad R_{2,t}=\{C_t<S_t\le k_t\},\quad
 R_{3,t}=\{p_t^{\mathrm{tr}}<S_t\le C_t\},\quad R_{4,t}=\{S_t\le p_t^{\mathrm{tr}}\},
\end{equation}
for the assumed ordering $p_t^{\mathrm{tr}}<C_t<k_t$. When this ordering fails, the partition must be revised to avoid overlap or empty incorrectly specified intervals. Equivalently the segmented design uses features $e_t\1_{R_{j,t}}$, $j=1,\ldots,4$, removing columns that are almost surely redundant. The region definition is evaluated at the decision time using the applicable conversion price and trigger levels. The segmented function space contains the global space if the same base features and region partition are used.

\subsection{Existence and uniqueness of least-squares coefficients}
For any fixed sample design $E\in\mathbb R^{M\times K}$ and response $y\in\mathbb R^M$, least-squares minimizers of $\norm{y-E\theta}_2^2$ always exist. They satisfy
\begin{equation}\label{eq:normalequation}
 E^{\top}E\theta=E^{\top}y.
\end{equation}
The coefficient vector is unique \emph{if and only if} $\operatorname{rank}(E)=K$. Only then is it correct to write $\theta=(E^{\top}E)^{-1}E^{\top}y$. A sample size $M>K$ alone does not imply full column rank. For rank-deficient designs the fitted vector $E\widehat\theta$ is still unique and the minimum-norm coefficient solution is $E^+y$, where $E^+$ is the pseudoinverse. Numerically, QR or SVD is preferable to explicitly forming $(E^{\top}E)^{-1}$.

\section{Convergence and approximation error}\label{sec:theory}
The proofs below adapt standard finite-dimensional regression and dynamic-programming arguments; they are included to specify the guarantees and limitations of the present construction \citep{Clement2002,Stentoft2004}. No general finite-sample rate or superiority of market-price fit is claimed.

\paragraph{Assumptions.} Fix a finite horizon $T$ and a feature design (global or segmented). (A1) The $M$ simulated complete histories are independent and identically distributed under the specified model and exogenous issuer policy, conditioned on the common initial state. (A2) Terminal and eligible exercise/call payments and the features are square-integrable, and the decision operators $D_t$ are $\F_t$-measurable with the form \eqref{eq:decision}. (A3) On the retained feature coordinates, $A_t=\E_{\Q}[e_te_t^{\top}]$ is finite and positive definite for each $t=1,\ldots,T-1$; zero-probability segmented regions are omitted. (A4) The initial state and immediate decision operator $D_0$ are fixed across replicated paths. These assumptions concern the \emph{simplified model} and cannot be inferred solely from the original empirical price tables.

\begin{lemma}[Non-expansiveness of the exercise decision]\label{lem:lipschitz}
For each fixed realized contract history, the map $D_t$ in \eqref{eq:decision} satisfies $|D_t(a)-D_t(b)|\le |a-b|$ for all $a,b\in\mathbb R$.
\end{lemma}
\begin{proof}
On a voluntary-exercise date the map $h\mapsto\max(G_t,h)$ is $1$-Lipschitz. Otherwise $D_t$ is either the identity or the constant forced-settlement map, both of which satisfy the same inequality.
\end{proof}

Define the population regression recursion, on the same full-history probability space, by
\begin{equation}\label{eq:population}
 \overline V_T=V_T,\qquad
 \theta_t=A_t^{-1}\E_{\Q}[e_t\beta\overline V_{t+1}],\qquad
 \overline V_t=D_t(e_t^{\top}\theta_t),\quad t=T-1,\ldots,1,
\end{equation}
with $\overline V_0=D_0(\E_{\Q}[\beta\overline V_1])$. The variable $\overline V_t$ is a population \emph{regression approximation}, not the exact model value $V_t$.

\begin{theorem}[Consistency for a fixed regression space]\label{thm:consistency}
Under (A1)--(A4), the estimator \eqref{eq:ols}--\eqref{eq:price} satisfies $\widehat V_0^{(M)}\longrightarrow\overline V_0$ in probability as $M\to\infty$, with fixed horizon and fixed feature dimension.
\end{theorem}
\begin{proof}
Proceed backwards. At time $T-1$, the regression response is the exact $\beta V_T$. By the weak law of large numbers and Cauchy--Schwarz integrability,
\[
 M^{-1}E_{T-1}^{\top}E_{T-1}\xrightarrow{P}A_{T-1},\qquad
 M^{-1}E_{T-1}^{\top}y_{T-1}\xrightarrow{P}\E[e_{T-1}\beta V_T].
\]
Because $A_{T-1}$ is positive definite, the empirical Gram matrix is invertible with probability tending to one, and continuity of inversion yields $\widehat\theta_{T-1}\xrightarrow{P}\theta_{T-1}$. The pseudoinverse rule on the exceptional singular event does not alter this limit.

Suppose $\widehat\theta_{t+1}\xrightarrow{P}\theta_{t+1}$. Evaluate $\overline V_{t+1}^{(i)}=D_{t+1}^{(i)}((e_{t+1}^{(i)})^{\top}\theta_{t+1})$ on each same simulated history. Lemma~\ref{lem:lipschitz} implies
\begin{align*}
 &\left\lVert\frac1M\sum_{i=1}^M e_t^{(i)}\beta\big(\widehat V_{t+1}^{(i)}-\overline V_{t+1}^{(i)}\big)\right\rVert\\
 &\qquad\le\beta\norm{\widehat\theta_{t+1}-\theta_{t+1}}\,\frac1M\sum_{i=1}^M\norm{e_t^{(i)}}\norm{e_{t+1}^{(i)}}=o_P(1).
\end{align*}
The final empirical average converges in probability to a finite expectation by the law of large numbers, again using Cauchy--Schwarz. After replacing the empirical responses by their population counterparts, the pathwise summands are i.i.d., so the ordinary law of large numbers and the same Gram-matrix argument give $\widehat\theta_t\xrightarrow{P}\theta_t$. Repeating reaches $t=1$. Finally, the difference between the sample means of $\beta\widehat V_1^{(i)}$ and $\beta\overline V_1^{(i)}$ is bounded by $\beta\norm{\widehat\theta_1-\theta_1}M^{-1}\sum_i\norm{e_1^{(i)}}=o_P(1)$. The remaining average converges to $\E[\beta\overline V_1]$. Continuity of $D_0$ (Lemma~\ref{lem:lipschitz}) completes the proof. For $T=1$ the claim is the ordinary law of large numbers for the terminal payoff.
\end{proof}

For $t=1,\ldots,T-1$, let $\HH_t=\operatorname{span}\{e_{t,1},\ldots,e_{t,K_t}\}\subset L^2(\F_t,\Q)$, let $\Pi_t$ denote the orthogonal $L^2(\Q)$ projection onto $\HH_t$, and write $\norm{X}_{2}=(\E_{\Q}|X|^2)^{1/2}$. Define the \emph{single-step exact-target} projection errors
\begin{equation}\label{eq:epsilon}
 \varepsilon_t=\norm{H_t-\Pi_t H_t}_2,\qquad H_t=\E_{\Q}[\beta V_{t+1}\mid\F_t].
\end{equation}

\begin{theorem}[Propagation of basis approximation error]\label{thm:error}
Under (A1)--(A4), the population recursion \eqref{eq:population} and exact recursion \eqref{eq:bellman} satisfy
\begin{equation}\label{eq:bound}
 |\overline V_0-V_0|\le\sum_{t=1}^{T-1}\beta^t\varepsilon_t.
\end{equation}
The right-hand side is zero for $T=1$. The initial state is fixed and its continuation expectation is evaluated without a finite-dimensional projection; with a nontrivial time-zero regression, an additional $\varepsilon_0$ term is required.
\end{theorem}
\begin{proof}
Because $e_t$ is $\F_t$-measurable, the tower property gives
$\Pi_t(\beta\overline V_{t+1})=\Pi_t\E[\beta\overline V_{t+1}\mid\F_t]$ and $\Pi_t(\beta V_{t+1})=\Pi_t H_t$. Using the $1$-Lipschitz property, the contraction of orthogonal projection in $L^2$, and the triangle inequality,
\begin{align*}
 \norm{\overline V_t-V_t}_2
 &\le\norm{\Pi_t(\beta\overline V_{t+1})-H_t}_2\\
 &\le\norm{\Pi_t\big(\beta(\overline V_{t+1}-V_{t+1})\big)}_2+\norm{\Pi_tH_t-H_t}_2\\
 &\le\beta\norm{\overline V_{t+1}-V_{t+1}}_2+\varepsilon_t.
\end{align*}
Both terminal values coincide. Iterating the recursion bounds $\norm{\overline V_1-V_1}_2$ by $\sum_{t=1}^{T-1}\beta^{t-1}\varepsilon_t$. At time zero, non-expansiveness of $D_0$, Jensen's inequality and Cauchy--Schwarz imply
$|\overline V_0-V_0|\le\beta\norm{\overline V_1-V_1}_2$, proving \eqref{eq:bound}.
\end{proof}

\begin{proposition}[One-step projection ordering for segmented regression]\label{prop:nested}
Suppose the four regions form a measurable partition and the same base features are used. For any \emph{fixed common} target $H\in L^2(\F_t,\Q)$, if $\HH_t^{\mathrm{global}}\subseteq\HH_t^{\mathrm{seg}}$, then
\begin{equation}\label{eq:nested}
 \norm{H-\Pi_t^{\mathrm{seg}}H}_2\le\norm{H-\Pi_t^{\mathrm{global}}H}_2.
\end{equation}
\end{proposition}
\begin{proof}
Every global linear combination $e_t^{\top}b$ can be expressed as $\sum_{j=1}^4\1_{R_{j,t}}e_t^{\top}b$, so the global space is contained in the segmented space. The distance from a fixed $H$ to a larger closed finite-dimensional subspace cannot increase.
\end{proof}

\begin{remark}[Limits of the projection comparison]
Proposition~\ref{prop:nested} compares projection errors for the \emph{same} target at a \emph{single} date. It does not establish that the two separately propagated population recursions have ordered final absolute errors. Nor does it account for the increase in estimation variance caused by fitting more regional coefficients on finite samples. In particular, it does not prove lower error against observed market prices, which also reflects model misspecification and issuer-policy assumptions. A quantitative finite-sample benefit would need additional assumptions and new analysis.
\end{remark}

\begin{remark}[What is, and is not, shown]
Theorem~\ref{thm:consistency} gives $\widehat V_0^{(M)}\to\overline V_0$, not in general convergence to the exact model value $V_0$. Theorem~\ref{thm:error} bounds the deterministic basis-approximation component. To obtain consistency to $V_0$, one must additionally control the projection error as the spaces grow and the finite-sample estimation error when their dimension changes with $M$. Neither is established here.
\end{remark}

\section{Empirical studies on CCB pricing}\label{sec:empirical}
\subsection{Market sample and units}
The original empirical study identified 487 outstanding CCBs on 18 January 2023, 30 rated AAA, and selected ten liquid AAA bonds. Reported daily valuations cover 118 trading dates from 18 January to 17 July 2023 using 5,000 paths per bond-date (1,180 bond-date valuations). These are the original sample specifications and \emph{have not been independently rerun} for this revision.

For implementation with annually quoted effective government-bond yield $u$, an internally consistent continuously compounded rate is $r=\log(1+u)$ and the per-trading-day discount factor is $\exp(-r/252)=(1+u)^{-1/252}$. If $s$ is the sample standard deviation of daily stock returns, annual volatility is estimated as $\sigma\approx\sqrt{252}\,s$. These conventions are to be checked against the original simulation and data code, particularly because equations in the earlier manuscript used different time units.

\begin{table}[htbp]\centering\caption{Selected Daqin convertible bond contract terms from the original empirical source (historical snapshot).}\begin{tabular}{ll}\toprule
Item & Reported value\\\midrule
Price & 120.480\\Conversion value & 121.22\\Conversion price & 6.22\\Maturity date & 2026-12-14\\Put payment & 100 plus accrued unpaid interest\\Maturity payment & 108.00\\Call/put/reset trigger levels & 130\% / 70\% / 85\% of conversion price\\Call / put observation rules & 15/30 and 30/30\\\bottomrule\end{tabular}\end{table}

\subsection{Original pricing comparisons}
For bond-level observations $i=1,\ldots,N$, let $d_i=(V_i^{\mathrm{mdl}}-V_i^{\mathrm{mkt}})/V_i^{\mathrm{mkt}}$. Define mean relative error (MRE), mean absolute relative error (MARE), and relative root mean square error (RMSE) by
\begin{equation}\label{eq:metrics}
 \mathrm{MRE}=N^{-1}\sum_i d_i,\quad\mathrm{MARE}=N^{-1}\sum_i|d_i|,\quad
 \mathrm{RMSE}=\big(N^{-1}\sum_i d_i^2\big)^{1/2}.
\end{equation}
The following historical tables give percentages, and each Mean row is the equal-weight average of the ten \emph{bond-level} statistics (not a pooled RMSE over all observations).

\begin{table}[htbp]\centering\caption{Original global-regression results (with both rolling-window features). Not recomputed.}\label{tab:global}\small
\begin{tabular}{lrrrr}\toprule Bond & Ticker & MRE (\%) & MARE (\%) & RMSE (\%)\\\midrule
Nan Hang & 110075 & -3.15 & 3.16 & 3.62\\
He Jian & 113024 & -6.24 & 6.24 & 6.26\\
Ji Dong & 127025 & -2.24 & 2.24 & 2.29\\
Chang Qi & 113049 & -4.78 & 4.78 & 5.18\\
Wen Shi & 123107 & -3.02 & 3.06 & 3.37\\
Zhong Te & 127056 & -3.67 & 3.67 & 3.95\\
Pu Fa & 110059 & -1.22 & 1.22 & 1.27\\
Xing Ye & 113052 & 0.07 & 0.55 & 0.80\\
Su Zu & 110083 & -0.07 & 2.44 & 3.02\\
Cai Tong & 113043 & -1.57 & 1.57 & 1.66\\\midrule
Mean & -- & -2.59 & 2.89 & 3.14\\\bottomrule\end{tabular}\end{table}

\begin{table}[htbp]\centering\caption{Original global-regression results excluding $Y_t$ and its interactions. Not recomputed.}\label{tab:noy}\small
\begin{tabular}{lrrrr}\toprule Bond & Ticker & MRE (\%) & MARE (\%) & RMSE (\%)\\\midrule
Nan Hang & 110075 & -3.83 & 3.84 & 4.22\\
He Jian & 113024 & -6.41 & 6.41 & 6.43\\
Ji Dong & 127025 & -2.25 & 2.25 & 2.29\\
Chang Qi & 113049 & -5.87 & 5.87 & 6.07\\
Wen Shi & 123107 & -3.83 & 3.83 & 4.02\\
Zhong Te & 127056 & -4.49 & 4.49 & 4.61\\
Pu Fa & 110059 & -1.54 & 1.54 & 1.68\\
Xing Ye & 113052 & 0.04 & 0.55 & 0.81\\
Su Zu & 110083 & -0.63 & 2.05 & 2.40\\
Cai Tong & 113043 & -1.57 & 1.57 & 1.66\\\midrule
Mean & -- & -3.04 & 3.24 & 3.42\\\bottomrule\end{tabular}\end{table}

\begin{table}[htbp]\centering\caption{Original four-region regression results (with both rolling-window features). Not recomputed.}\label{tab:segmented}\small
\begin{tabular}{lrrrr}\toprule Bond & Ticker & MRE (\%) & MARE (\%) & RMSE (\%)\\\midrule
Nan Hang & 110075 & -2.45 & 2.50 & 2.93\\
He Jian & 113024 & -5.77 & 5.77 & 5.83\\
Ji Dong & 127025 & -2.25 & 2.25 & 2.29\\
Chang Qi & 113049 & -5.13 & 5.13 & 5.37\\
Wen Shi & 123107 & -1.66 & 2.02 & 2.33\\
Zhong Te & 127056 & -3.72 & 3.72 & 3.98\\
Pu Fa & 110059 & -1.52 & 1.52 & 1.66\\
Xing Ye & 113052 & 0.04 & 0.56 & 0.82\\
Su Zu & 110083 & 0.77 & 2.14 & 2.73\\
Cai Tong & 113043 & -1.57 & 1.57 & 1.66\\\midrule
Mean & -- & -2.32 & 2.72 & 2.96\\\bottomrule\end{tabular}\end{table}

The original mean bond-level relative RMSE decreases from 3.42\% to 3.14\% when the $Y_t$-related features are included, and to 2.96\% after four-region segmentation. These are \emph{market-discrepancy} statistics, not numerical errors relative to an exact price. Improvements do not hold for every bond (e.g., the segmented Pu Fa RMSE is higher), and they do not establish the assumptions of Section~\ref{sec:theory}. The original study also included plots of model and market prices; original assets may be restored without changing the numerical tables.

\begin{figure}[p]\centering
\begin{subfigure}{0.46\textwidth}\centering
\includegraphics[width=\linewidth,height=0.115\textheight,keepaspectratio]{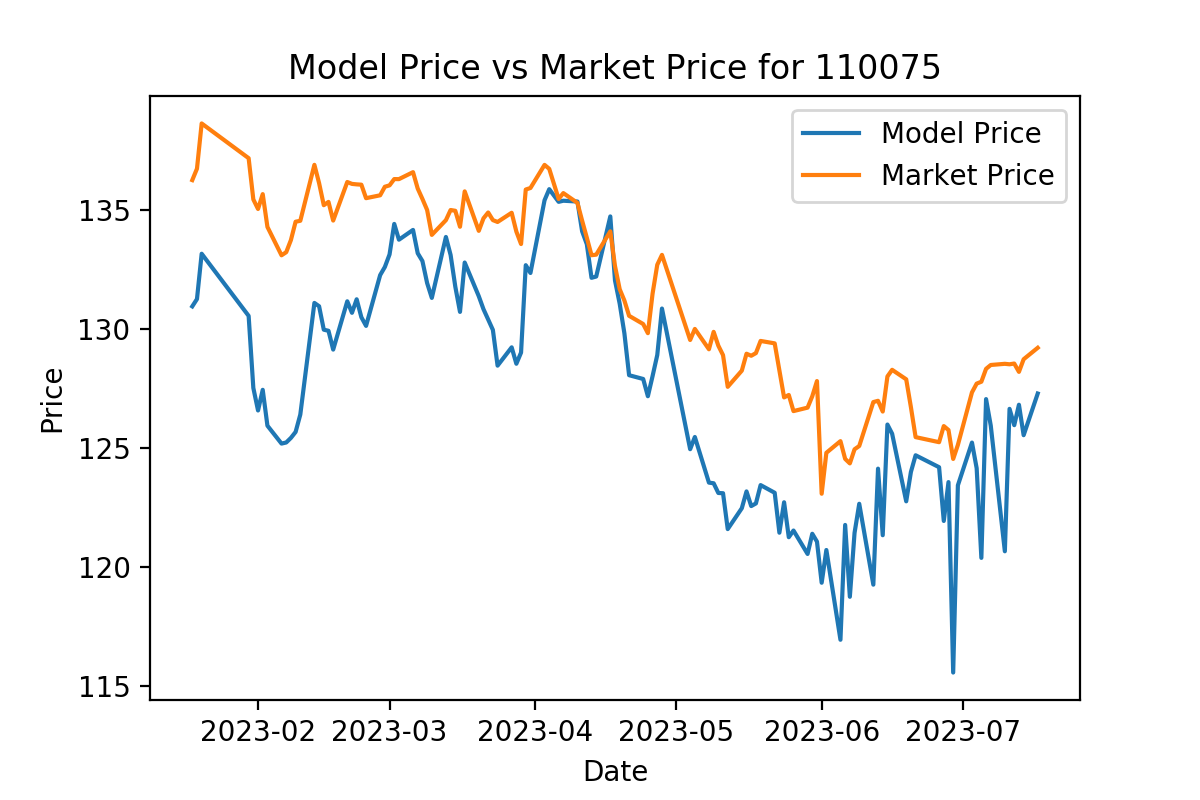}
\caption{Nan Hang}\end{subfigure}
\hfill
\begin{subfigure}{0.46\textwidth}\centering
\includegraphics[width=\linewidth,height=0.115\textheight,keepaspectratio]{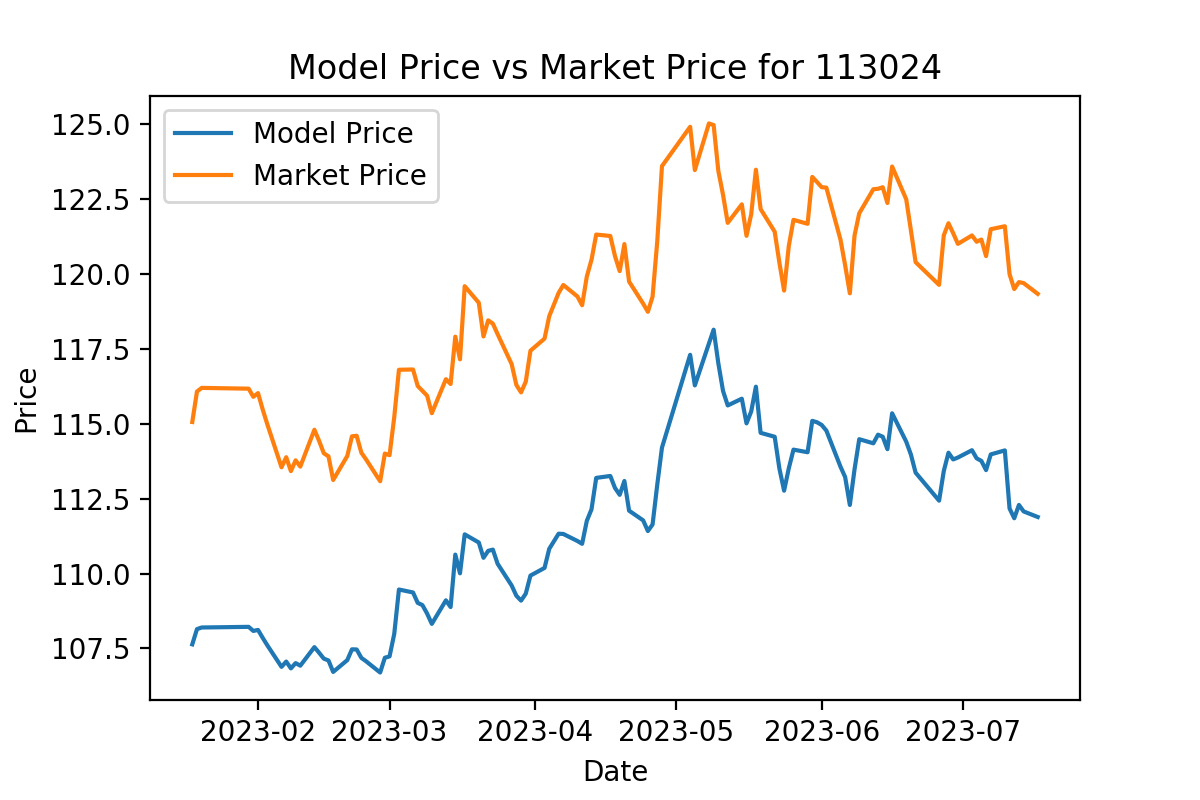}
\caption{He Jian}\end{subfigure}
\par\smallskip
\begin{subfigure}{0.46\textwidth}\centering
\includegraphics[width=\linewidth,height=0.115\textheight,keepaspectratio]{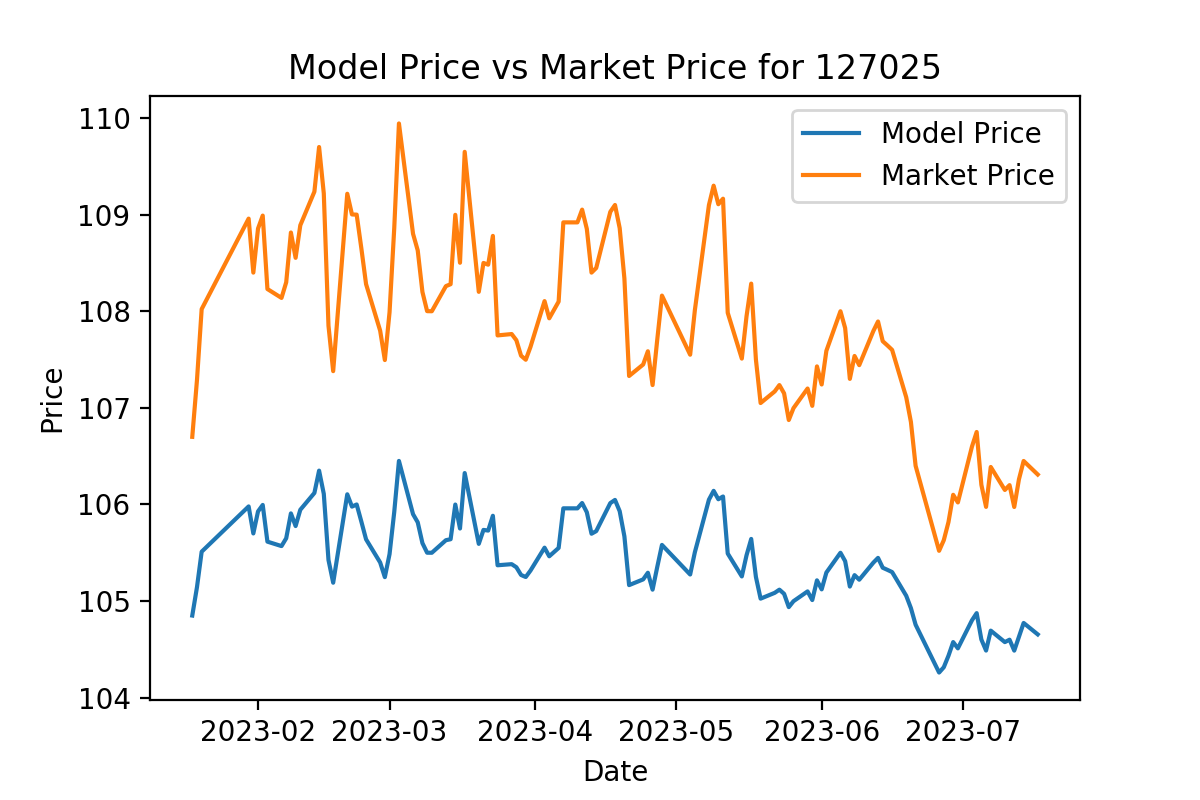}
\caption{Ji Dong}\end{subfigure}
\hfill
\begin{subfigure}{0.46\textwidth}\centering
\includegraphics[width=\linewidth,height=0.115\textheight,keepaspectratio]{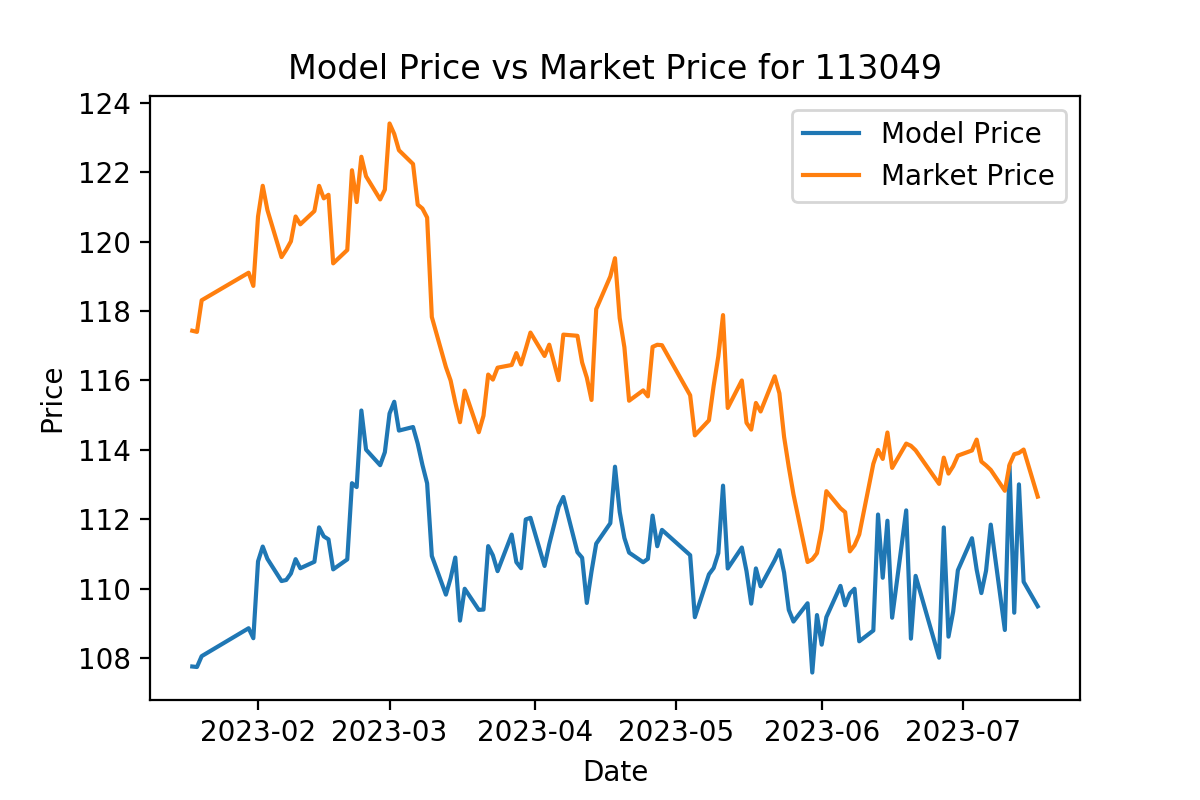}
\caption{Chang Qi}\end{subfigure}
\par\smallskip
\begin{subfigure}{0.46\textwidth}\centering
\includegraphics[width=\linewidth,height=0.115\textheight,keepaspectratio]{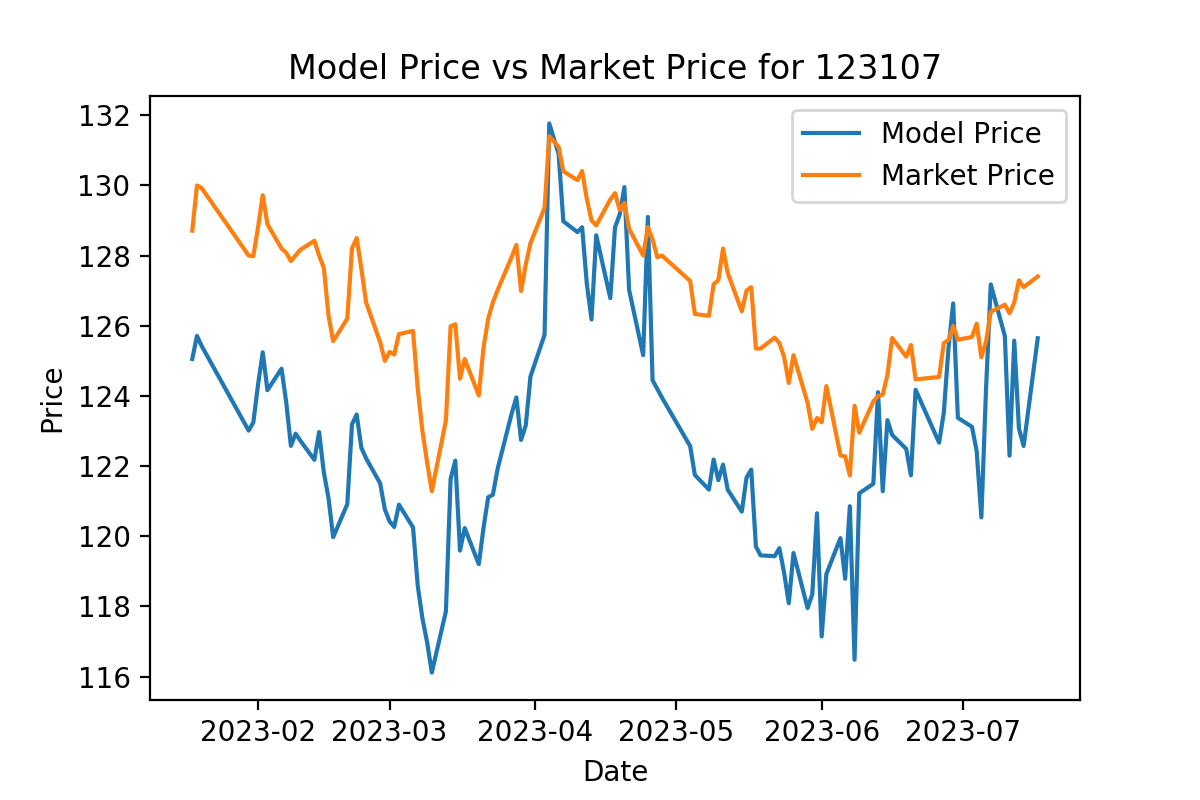}
\caption{Wen Shi}\end{subfigure}
\hfill
\begin{subfigure}{0.46\textwidth}\centering
\includegraphics[width=\linewidth,height=0.115\textheight,keepaspectratio]{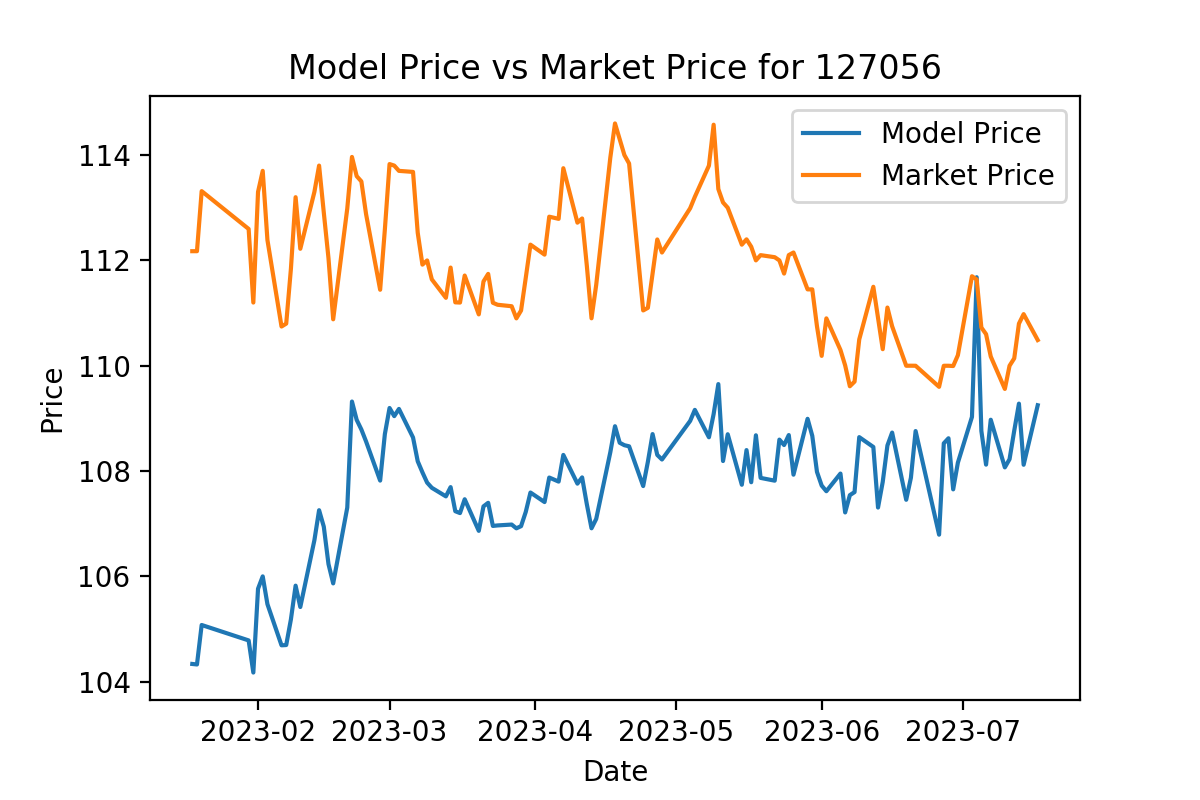}
\caption{Zhong Te}\end{subfigure}
\par\smallskip
\begin{subfigure}{0.46\textwidth}\centering
\includegraphics[width=\linewidth,height=0.115\textheight,keepaspectratio]{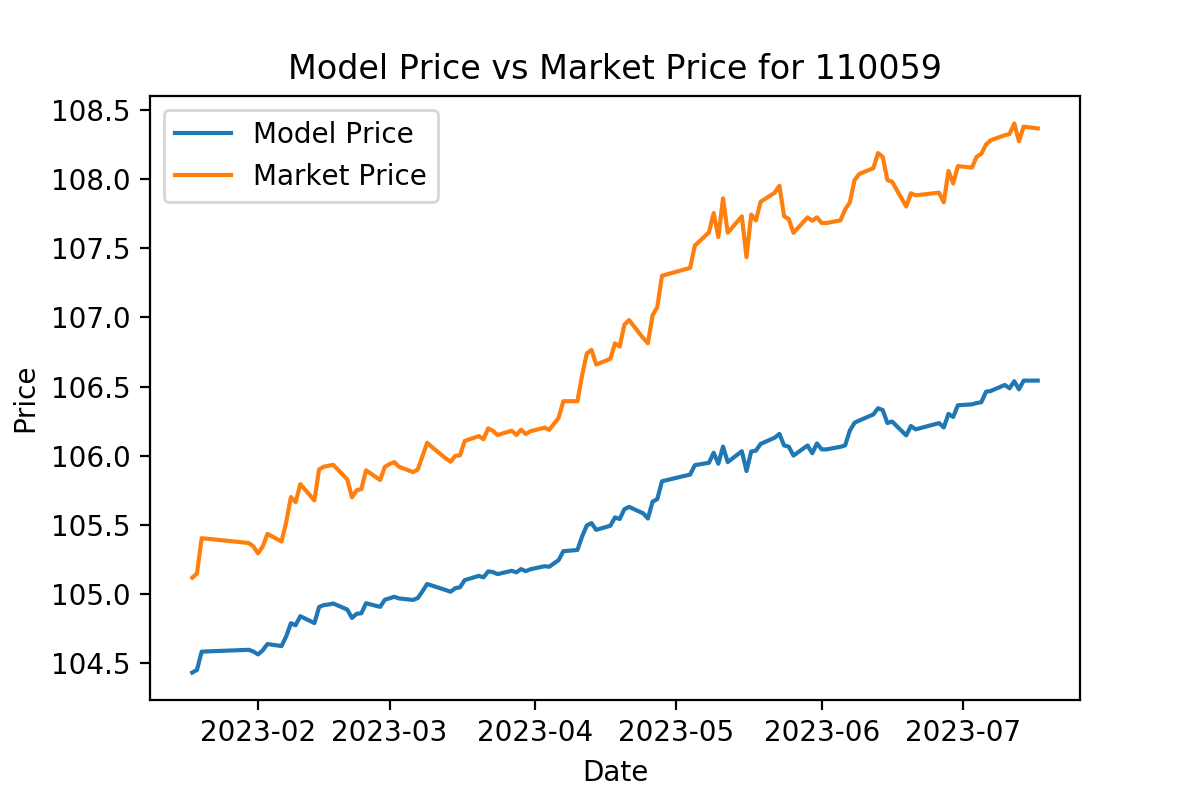}
\caption{Pu Fa}\end{subfigure}
\hfill
\begin{subfigure}{0.46\textwidth}\centering
\includegraphics[width=\linewidth,height=0.115\textheight,keepaspectratio]{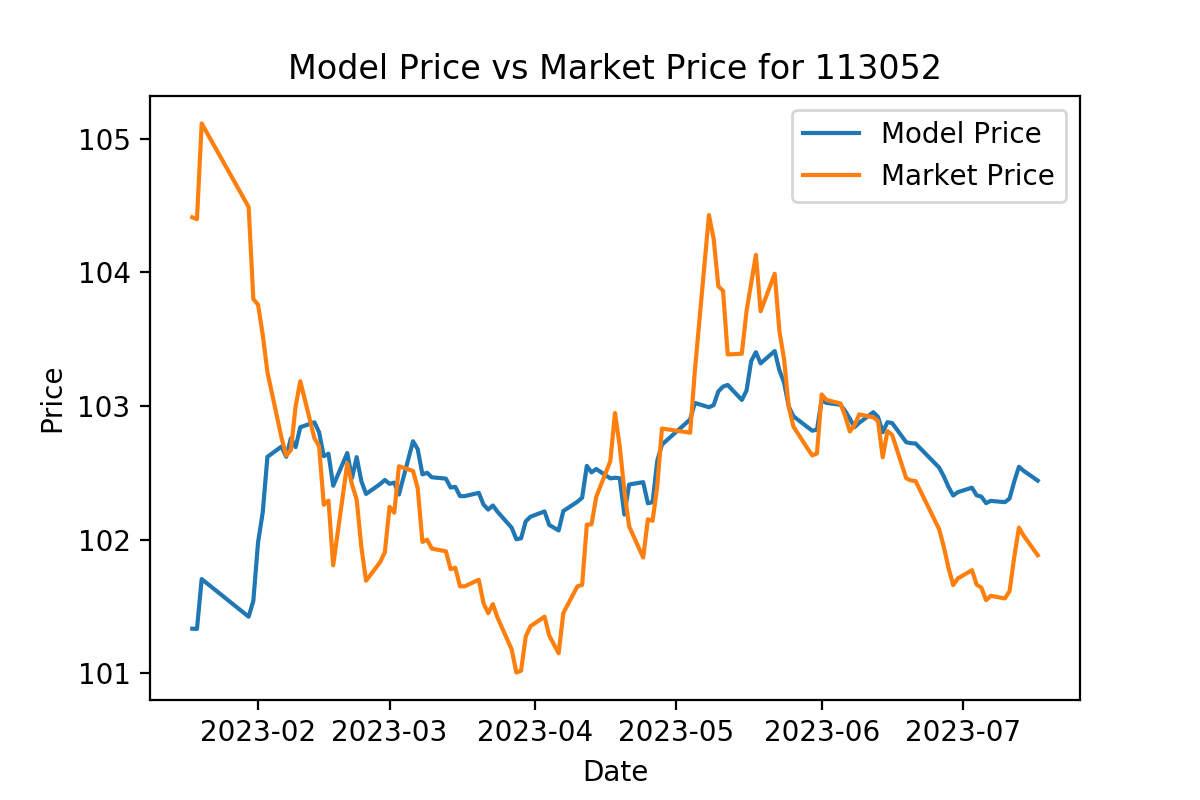}
\caption{Xing Ye}\end{subfigure}
\par\smallskip
\begin{subfigure}{0.46\textwidth}\centering
\includegraphics[width=\linewidth,height=0.115\textheight,keepaspectratio]{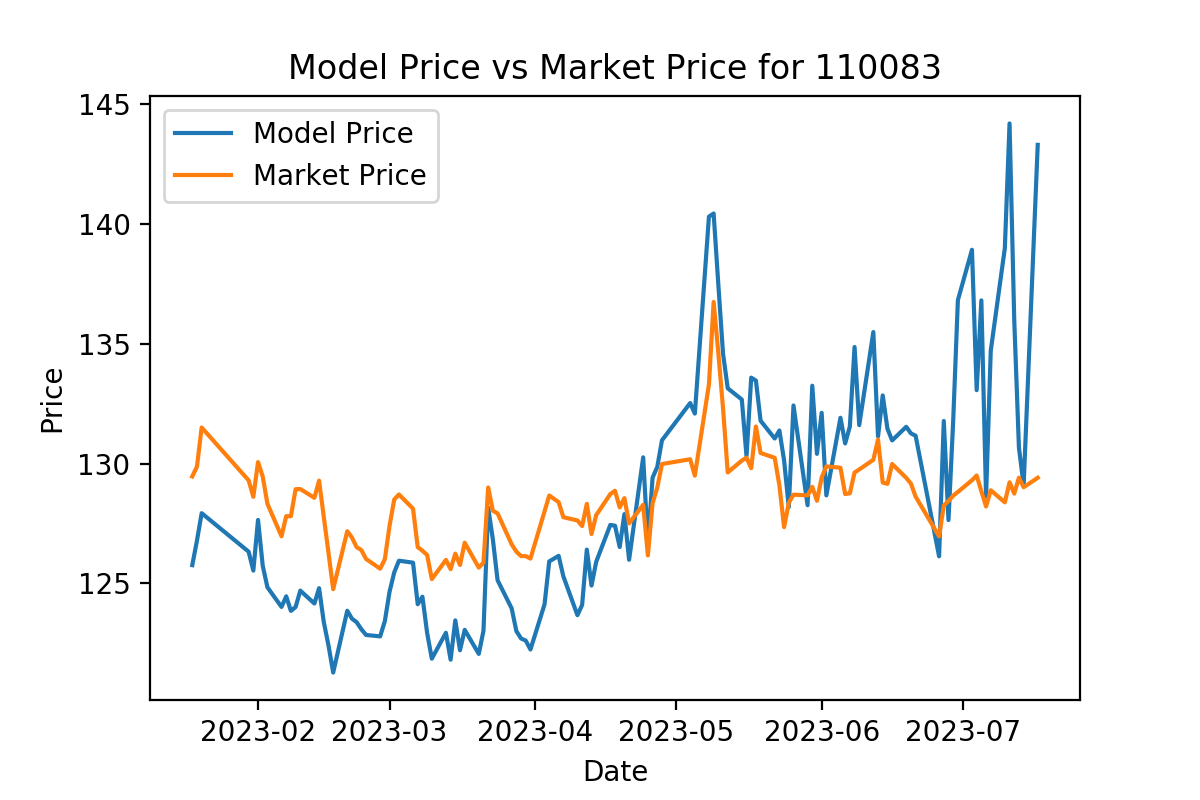}
\caption{Su Zu}\end{subfigure}
\hfill
\begin{subfigure}{0.46\textwidth}\centering
\includegraphics[width=\linewidth,height=0.115\textheight,keepaspectratio]{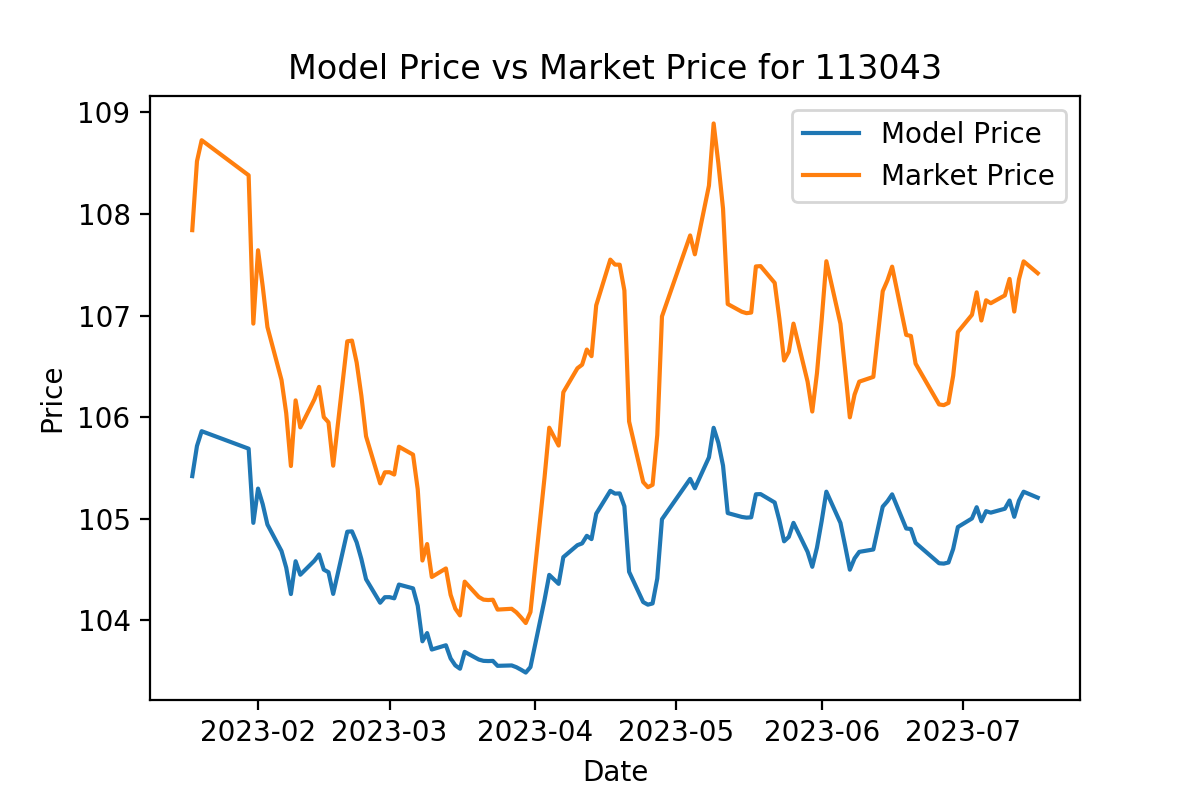}
\caption{Cai Tong}\end{subfigure}
\par\smallskip
\caption{Original global-regression model and market price plots for the ten CCBs (extracted from the historical arXiv PDF).}
\label{fig:global-ten}
\end{figure}
\begin{figure}[p]\centering
\begin{subfigure}{0.46\textwidth}\centering
\includegraphics[width=\linewidth,height=0.115\textheight,keepaspectratio]{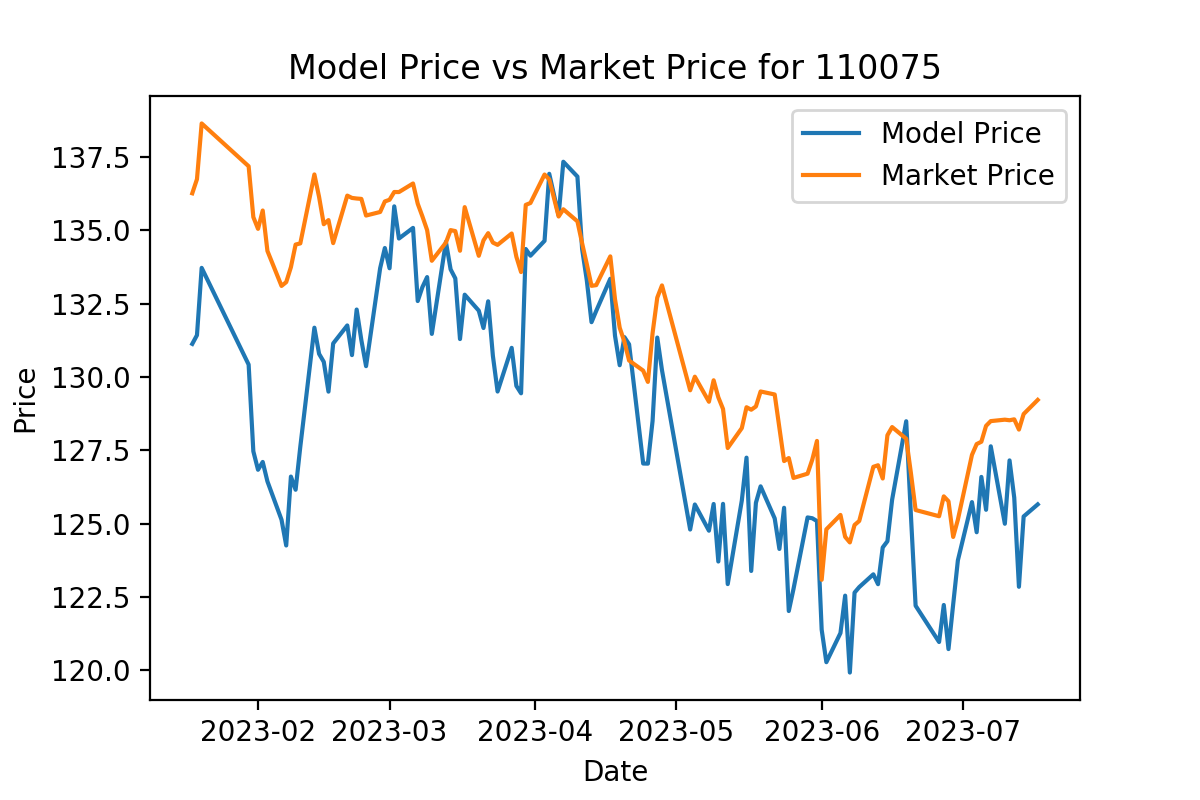}
\caption{Nan Hang}\end{subfigure}
\hfill
\begin{subfigure}{0.46\textwidth}\centering
\includegraphics[width=\linewidth,height=0.115\textheight,keepaspectratio]{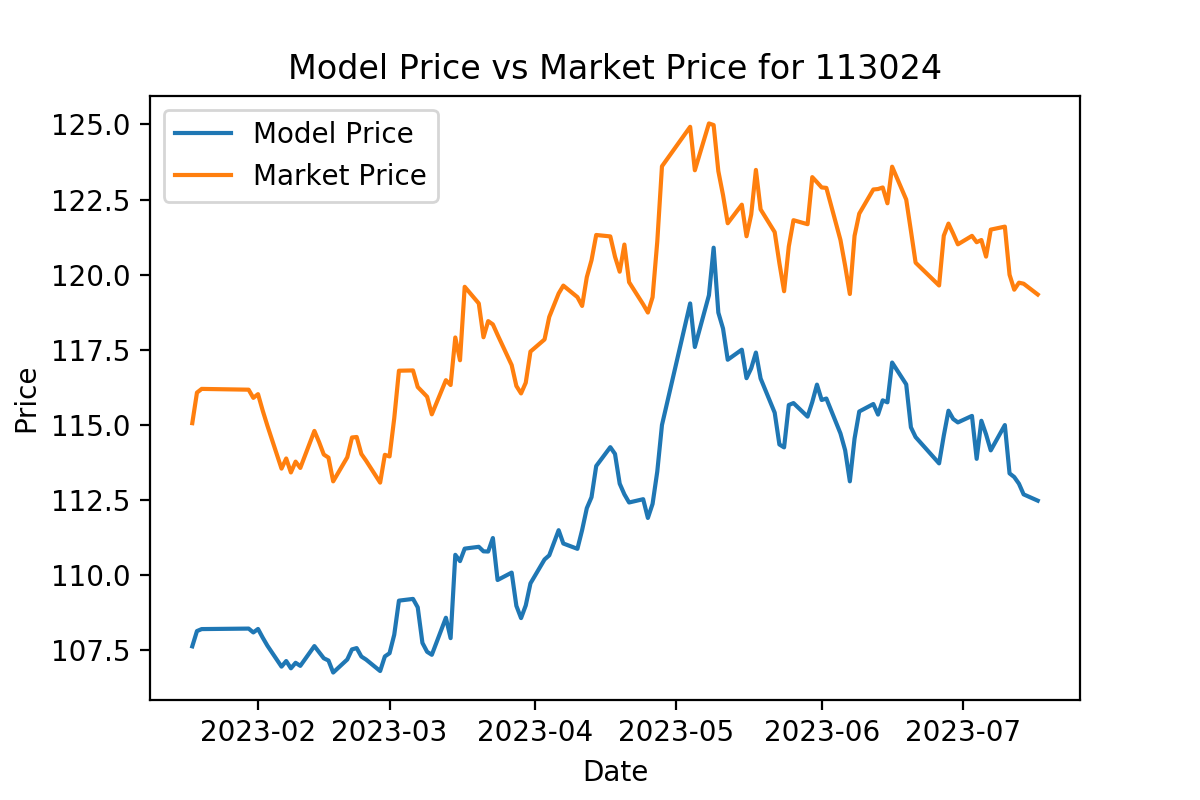}
\caption{He Jian}\end{subfigure}
\par\smallskip
\begin{subfigure}{0.46\textwidth}\centering
\includegraphics[width=\linewidth,height=0.115\textheight,keepaspectratio]{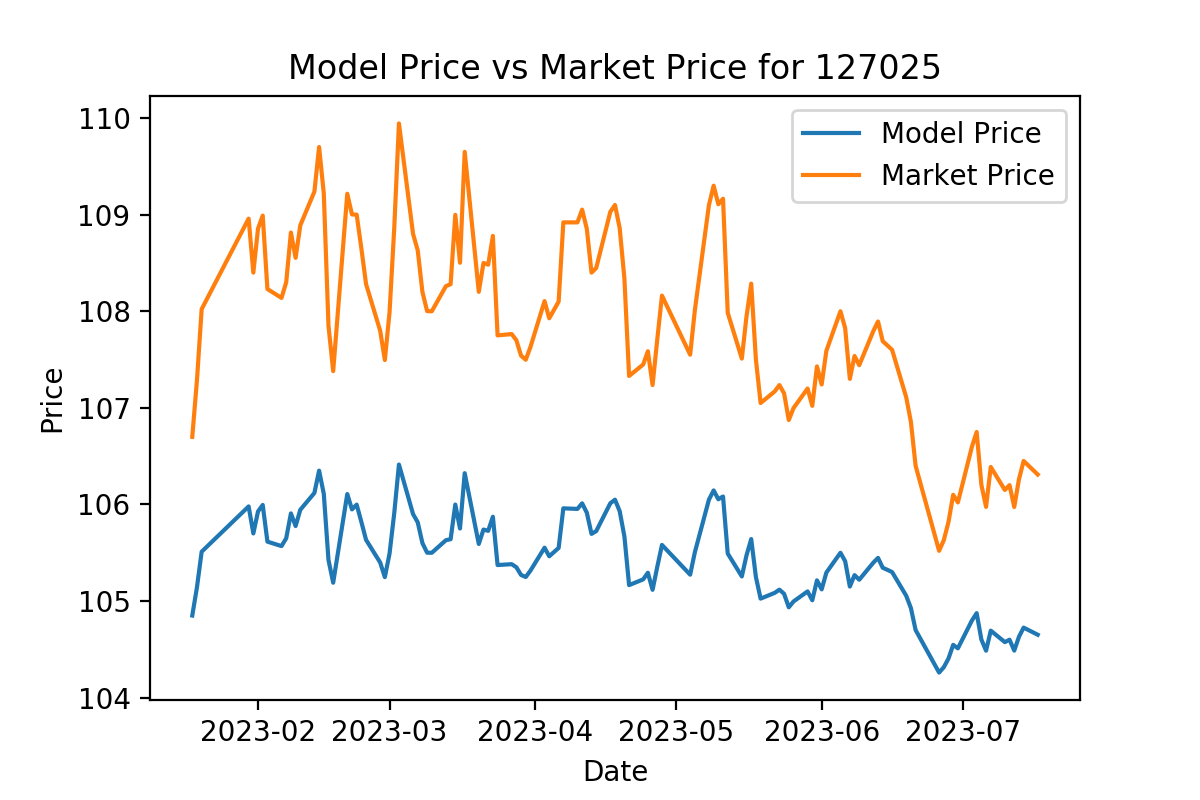}
\caption{Ji Dong}\end{subfigure}
\hfill
\begin{subfigure}{0.46\textwidth}\centering
\includegraphics[width=\linewidth,height=0.115\textheight,keepaspectratio]{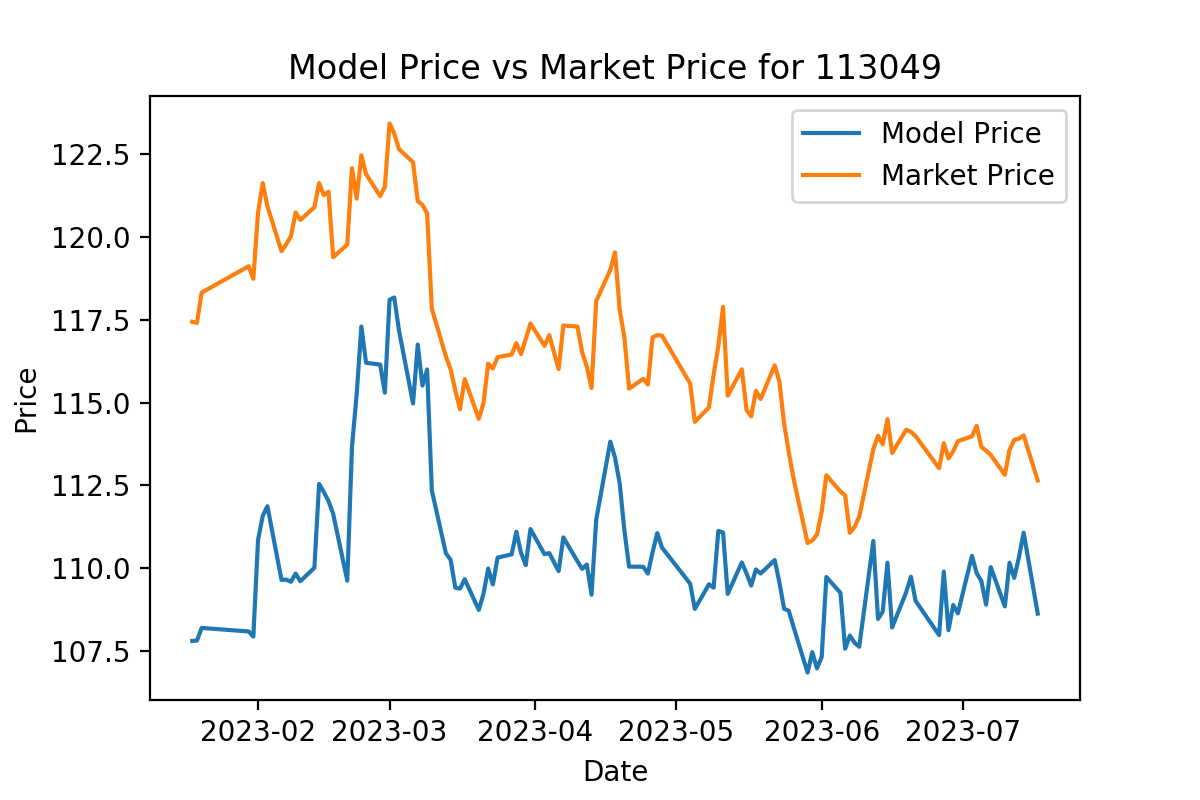}
\caption{Chang Qi}\end{subfigure}
\par\smallskip
\begin{subfigure}{0.46\textwidth}\centering
\includegraphics[width=\linewidth,height=0.115\textheight,keepaspectratio]{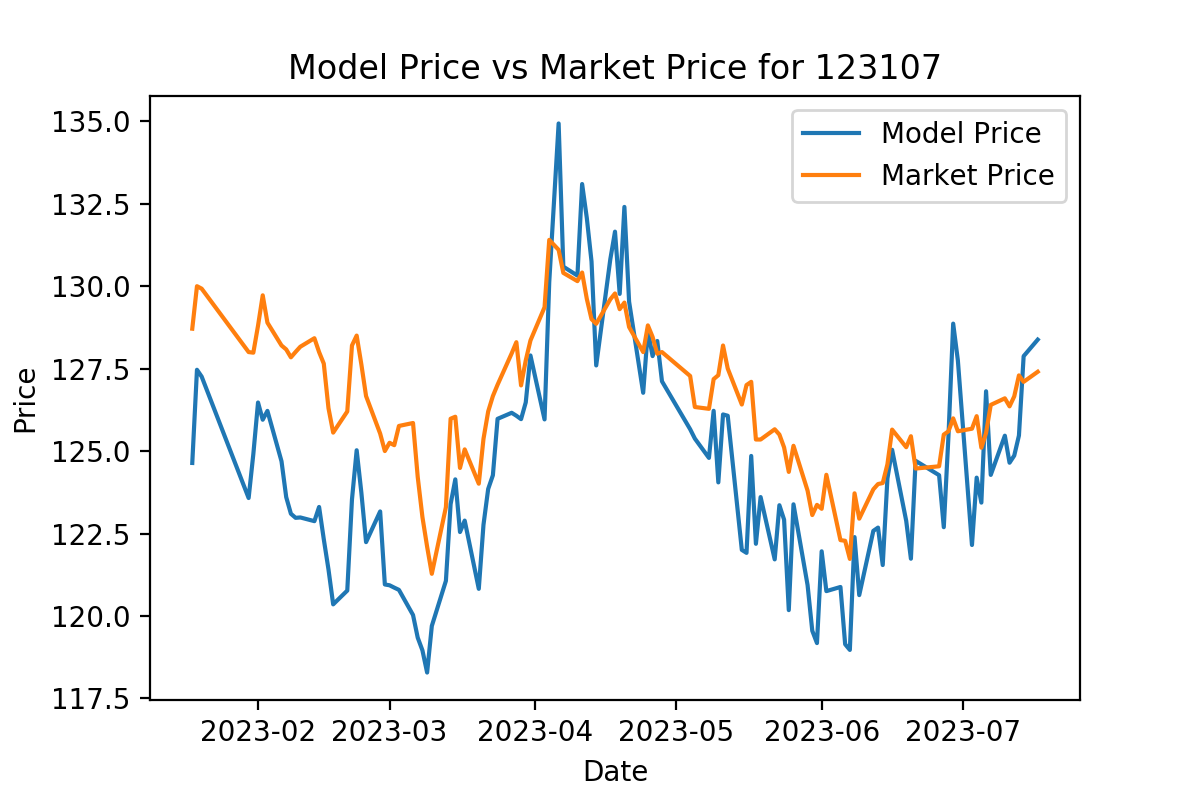}
\caption{Wen Shi}\end{subfigure}
\hfill
\begin{subfigure}{0.46\textwidth}\centering
\includegraphics[width=\linewidth,height=0.115\textheight,keepaspectratio]{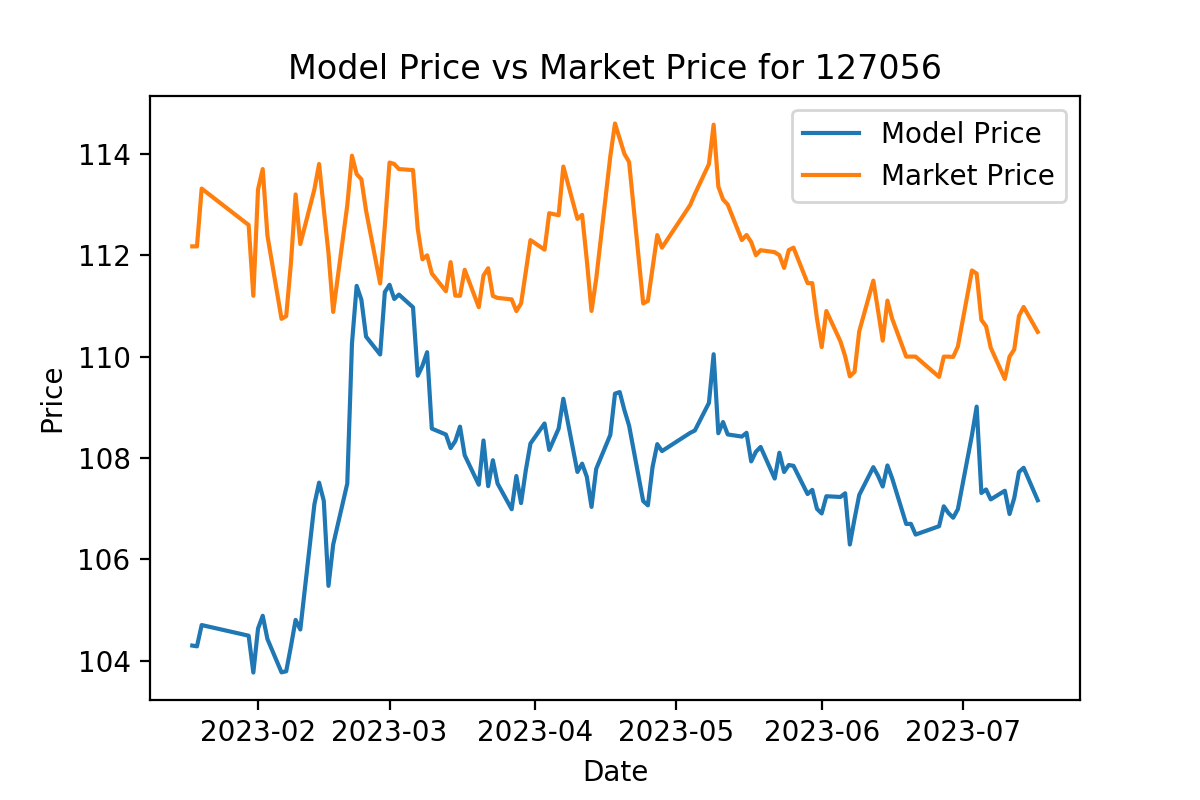}
\caption{Zhong Te}\end{subfigure}
\par\smallskip
\begin{subfigure}{0.46\textwidth}\centering
\includegraphics[width=\linewidth,height=0.115\textheight,keepaspectratio]{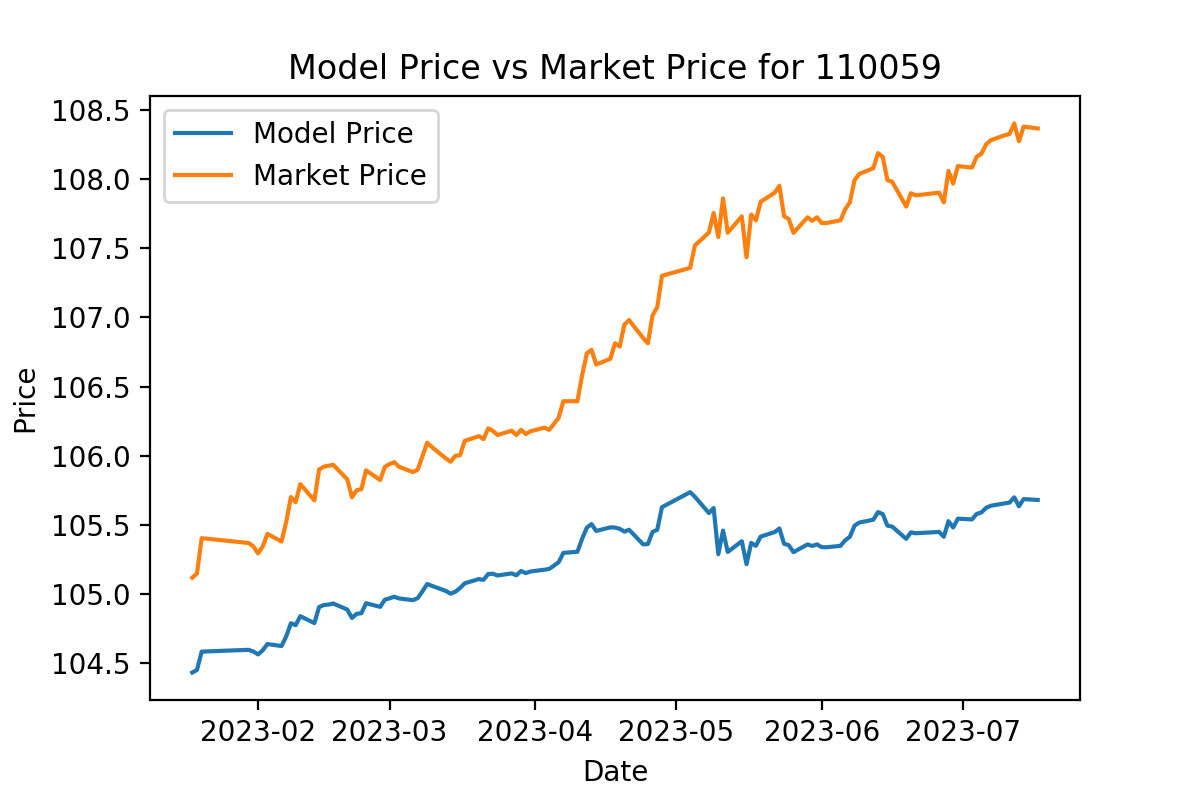}
\caption{Pu Fa}\end{subfigure}
\hfill
\begin{subfigure}{0.46\textwidth}\centering
\includegraphics[width=\linewidth,height=0.115\textheight,keepaspectratio]{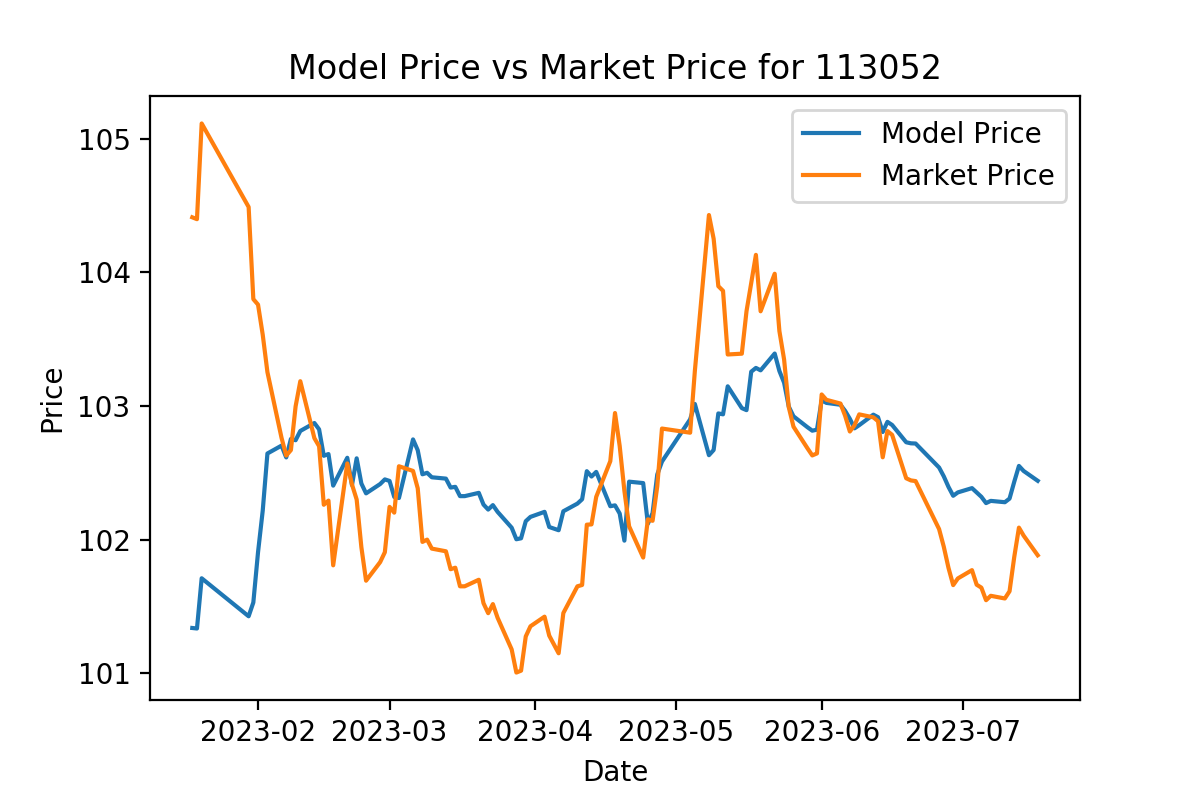}
\caption{Xing Ye}\end{subfigure}
\par\smallskip
\begin{subfigure}{0.46\textwidth}\centering
\includegraphics[width=\linewidth,height=0.115\textheight,keepaspectratio]{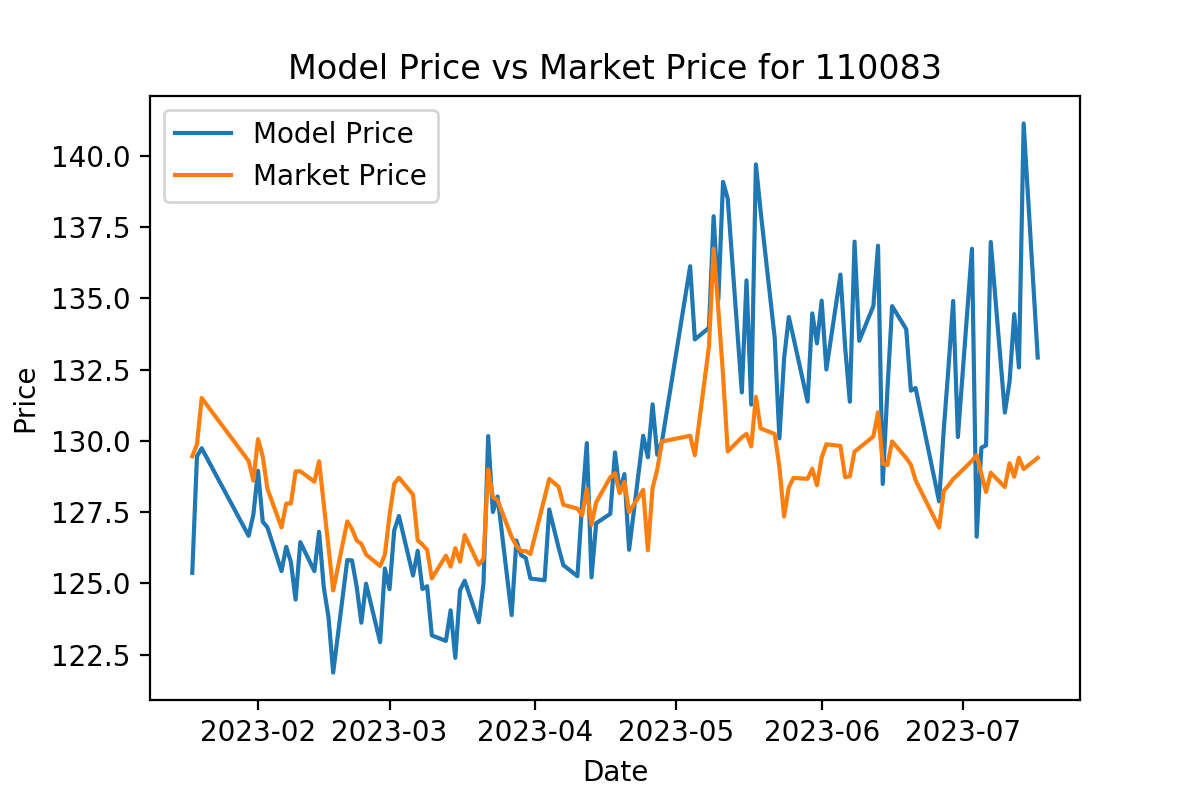}
\caption{Su Zu}\end{subfigure}
\hfill
\begin{subfigure}{0.46\textwidth}\centering
\includegraphics[width=\linewidth,height=0.115\textheight,keepaspectratio]{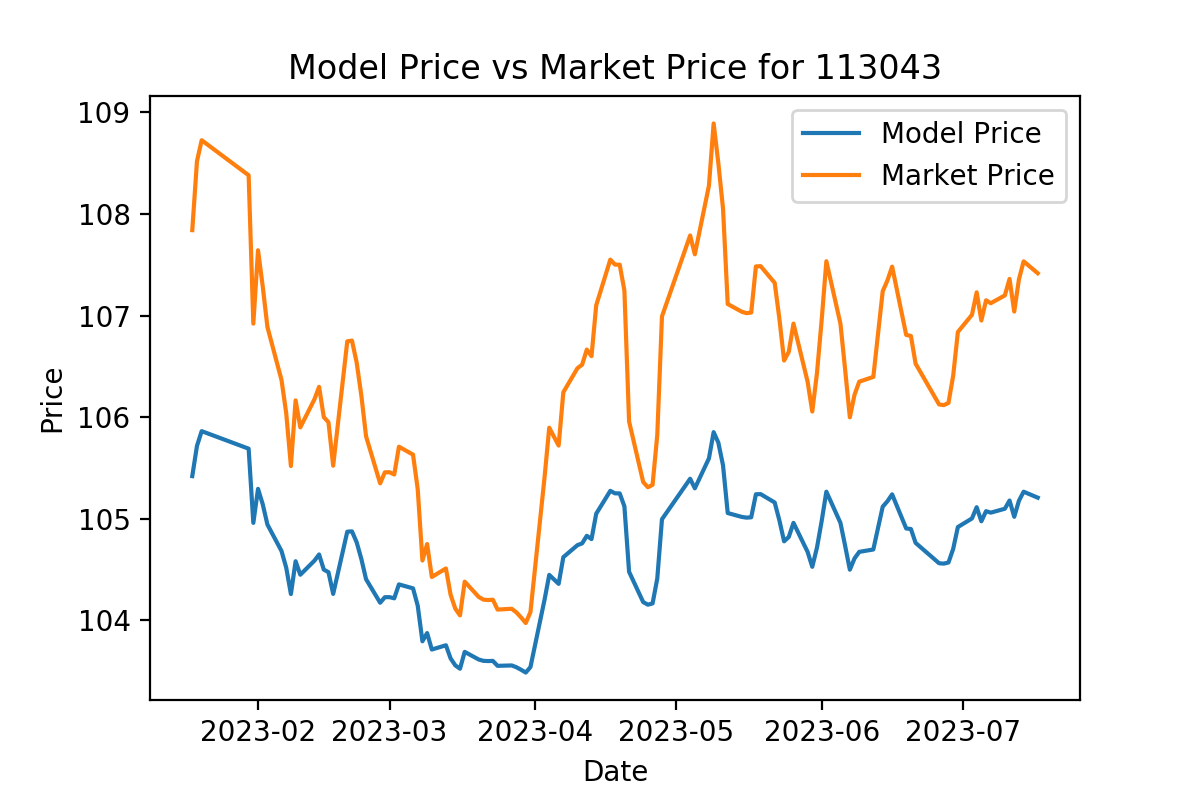}
\caption{Cai Tong}\end{subfigure}
\par\smallskip
\caption{Original segmented-regression model and market price plots for the ten CCBs (extracted from the historical arXiv PDF).}
\label{fig:segmented-ten}
\end{figure}

\subsection{Independent numerical benchmark under a matched simplified contract}\label{sec:benchmark}
We provide an independently reproducible test of the numerical recursion using a simplified short-horizon \emph{discrete} contract. This experiment is intentionally distinct from the historical GBM-based CCB market comparisons. Both the exhaustive dynamic-programming reference and the Monte Carlo estimator use the \emph{same} Cox--Ross--Rubinstein (CRR) transition probabilities and the same contract conventions. Consequently the discrepancy below is not contaminated by comparing prices under different underlying-stock models.

In the test, $S_0=C_0=FV=100$, $B=105$, the immediate call settlement is $103$ and the put payment is $101$. We take $r=0.025$, $q=0$, $\sigma=0.30$, nine periods of $\Delta t=1/12$ year, a three-observation trigger window, and a two-out-of-three trigger rule. The upward and downward CRR stock multipliers and risk-neutral up probability are
\[
 u=\exp(\sigma\sqrt{\Delta t}),\qquad d=u^{-1},\qquad
 \pi=\frac{\exp((r-q)\Delta t)-d}{u-d}.
\]
Conversion is eligible from date 1, puts and resets from date 2, and calls from date 3. The test call and put thresholds are $1.25C_t$ and $0.93C_t$, evaluated immediately \emph{before} a possible same-date reset. Every qualifying put trigger causes a deterministic reset, when permitted, to the lower of the current conversion price and the greater of the current price and the last three observations' average. Reset decisions precede holder exercise decisions. Triggers are evaluated using the current pre-reset conversion price for the whole short observation window; this is a \emph{benchmark-only convention}, not a claim about the exact historical observation rules of all Chinese CCBs. The simplified model uses no intermediate coupons, credit losses or fees.

The reference algorithm enumerates complete up/down price histories and evaluates the backward recursion exactly on this finite tree; a single stock-price node is \emph{not} assumed to determine the value, because triggers and resets are path dependent. It yields the model reference value
\[
 V^{\mathrm{ref}}_0=114.223194\quad\text{(six decimal places)}.
\]
For the Monte Carlo computation the regression feature vector is $(1,z,z^2,F,F^2,Y,Y^2,zF,zY,FY)$ with $z=S/100$; its intercept and scaling differ from the original historical nine-feature specification, and are fixed in advance for numerical conditioning. The segmented design uses the four regions defined by the test contract and fits each region separately. Both algorithms use the projected-value recursion in \eqref{eq:ols}, a least-squares pseudoinverse, and identical simulated paths for a paired comparison within each replicate. We use 60 independent random seeds per $M\in\{1000,5000,10000,50000\}$.

For each method and sample size, write $\widehat V_0^{(m)}$ for the output from replicate $m$, $m=1,\ldots,R$ with $R=60$. We report
\begin{align*}
 \mathrm{Bias}&=R^{-1}\sum_{m=1}^{R}(\widehat V_0^{(m)}-V_0^{\mathrm{ref}}),\\
 \mathrm{SD}&=\left[\frac{1}{R-1}\sum_{m=1}^{R}(\widehat V_0^{(m)}-\overline{\widehat V}_0)^2\right]^{1/2},\\
 \mathrm{RMSE}&=\left[\frac{1}{R}\sum_{m=1}^{R}(\widehat V_0^{(m)}-V_0^{\mathrm{ref}})^2\right]^{1/2}.
\end{align*}
\begin{table}[htbp]\centering\caption{Independent CRR-model benchmark against the exhaustive dynamic-programming reference $114.223194$. Sixty independent replications per sample size; all entries are in price units rather than percentages.}\label{tab:benchmark}
\begin{tabular}{llrrrr}\toprule Method & $M$ & Mean price & Bias & SD & RMSE\\\midrule
Global & 1,000 & 115.1450 & +0.9218 & 0.3947 & 1.0015\\
Four-region & 1,000 & 114.8193 & +0.5961 & 0.3967 & 0.7142\\
Global & 5,000 & 115.1901 & +0.9669 & 0.1572 & 0.9794\\
Four-region & 5,000 & 114.8137 & +0.5905 & 0.1581 & 0.6110\\
Global & 10,000 & 115.1834 & +0.9602 & 0.1200 & 0.9676\\
Four-region & 10,000 & 114.8045 & +0.5813 & 0.1235 & 0.5940\\
Global & 50,000 & 115.2036 & +0.9804 & 0.0549 & 0.9819\\
Four-region & 50,000 & 114.8133 & +0.5901 & 0.0533 & 0.5924\\
\bottomrule\end{tabular}\end{table}
\begin{figure}[htbp]\centering
 \includegraphics[width=0.80\linewidth]{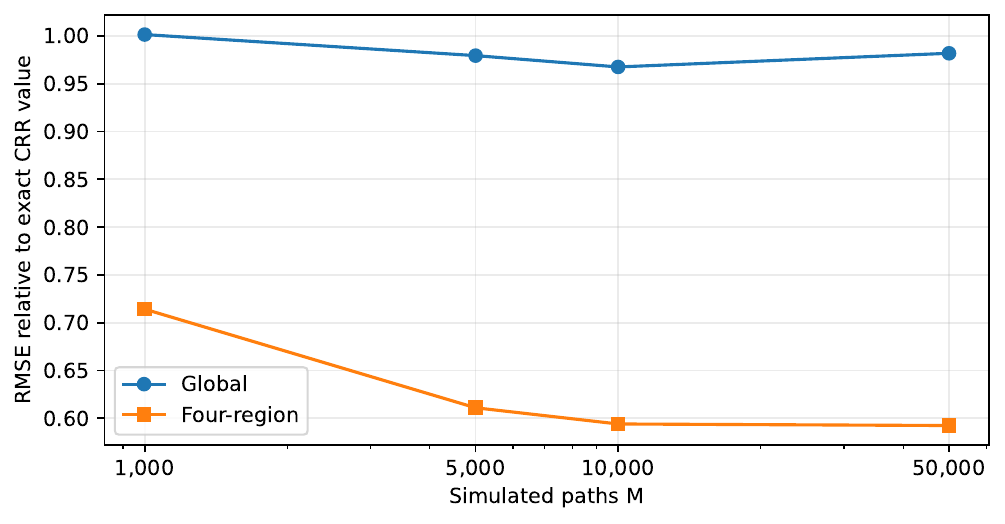}
 \caption{RMSE of the global and four-region regression algorithms relative to the exactly evaluated, matching nine-period CRR model. Each point summarizes 60 independent replications; connecting lines guide the eye.}\label{fig:benchmark_rmse}
\end{figure}
In this particular simplified model, the segmented approach has lower RMSE than the global regression for every tested sample size. Increasing $M$ reduces observed replication variability, but the error relative to the exact model value remains nonzero for both fixed feature spaces. This observation is consistent with Theorem~\ref{thm:consistency}, which concerns convergence to a \emph{population regression approximation}, not necessarily to the exact value. The experiment does not establish a theoretical rate, numerical convergence for every CCB contract, or superiority for arbitrary partitions. It also does not measure error against the exact population regression limit $\overline V_0$: that would require a separate population-approximation reference. The code and per-replication values are supplied as supplementary reproducibility materials.

\section{Trading strategy based on the model prices}\label{sec:trading}
The original study ranked outstanding bonds by model-implied undervaluation, held ten positions and rebalanced daily over the reported February--July 2023 sample, assuming a 0.1\% transaction-cost convention. The following results are carried over unchanged; the reported fee convention and portfolio construction have not been independently audited.
\begin{table}[htbp]\centering\caption{Original backtest metrics, retained without recalculation.}\begin{tabular}{lrrr}\toprule
Method & Cumulative return (\%) & Sharpe ratio & Maximum drawdown (\%)\\\midrule
Global least squares & 21.94 & 2.09 & 7.76\\
Four-region regression & 29.17 & 1.20 & 20.00\\
Double-low benchmark & 3.55 & 0.38 & 23.89\\\bottomrule\end{tabular}\end{table}
\pathfigure{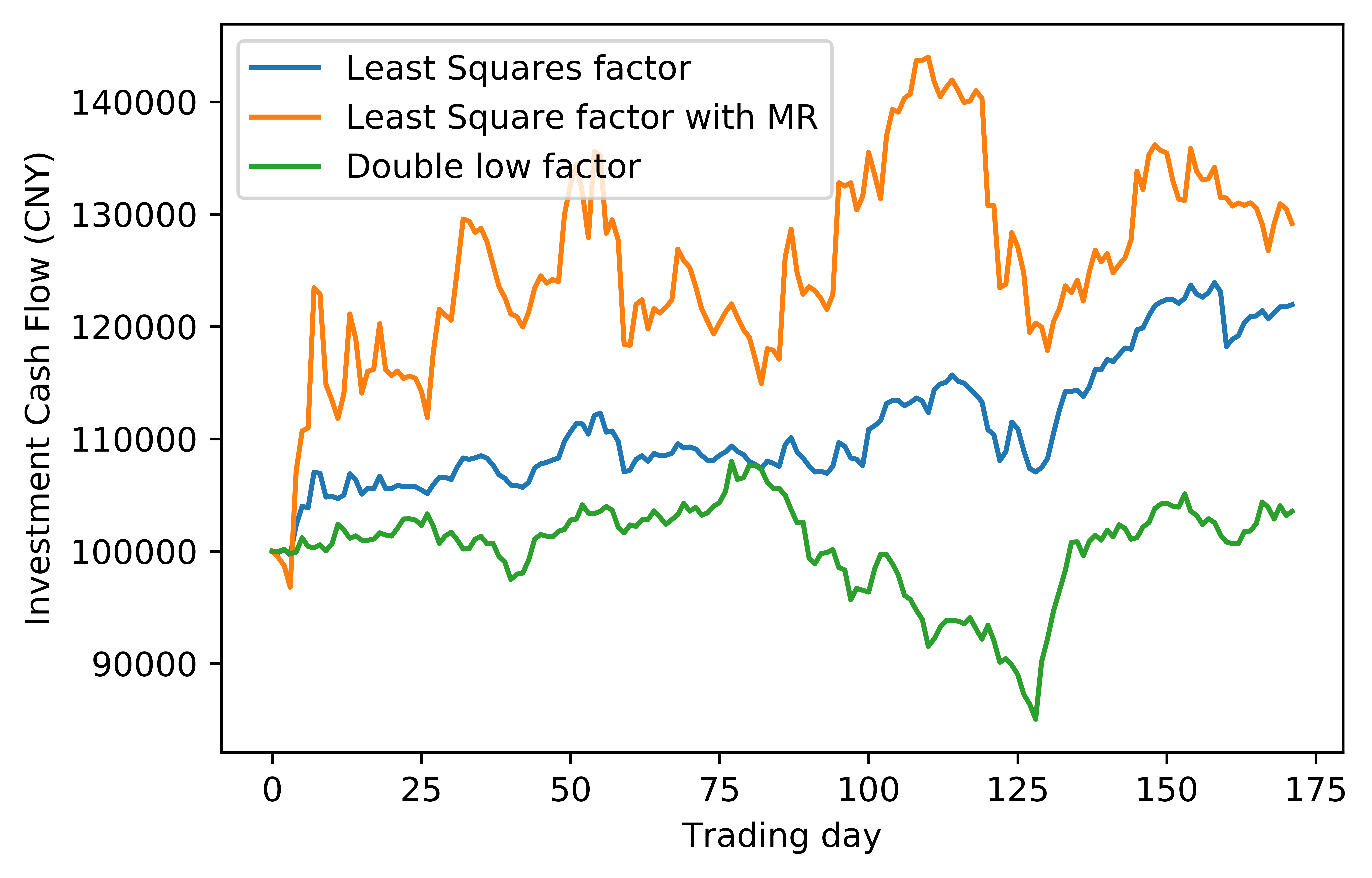}{0.77}{Original cumulative trading performance plot, if source graphic is supplied.}
The segmented portfolio produced a higher cumulative return in this sample, but its maximum drawdown rose and its Sharpe ratio fell relative to the global-regression portfolio. These historical returns are not evidence of a guaranteed out-of-sample trading advantage.

\section{Discussion and limitations}
The original contribution is an application-specific CCB valuation framework that uses contractual rolling-window features and segmented regression, with an exogenous stochastic reset assumption. The theory now makes explicit which features of the algorithm admit elementary finite-dimensional guarantees. Its statistical convergence is to a \emph{population projected recursion}; the basis approximation error is bounded by single-step projections. The segmented space has a smaller or equal projection error for a common target, but finite-sample estimates can be noisier and their final pricing errors need not be ordered.

Three limitations require particular care. First, the simplified issuer policy does not solve a discretionary issuer--holder stochastic game; a state-dependent empirical reset model may be preferable to a fixed probability. Second, the correct GBM drift and annual/daily scaling must be reconciled with the original code, and affected empirical tables rerun if the implementation was inconsistent. Third, comparisons to observed market quotations conflate numerical approximation with omitted coupons, credit/liquidity effects, parameter uncertainty and market-pricing deviations. A model-consistent numerical benchmark is reported in Section~\ref{sec:benchmark}; it supplies evidence about this simplified contract, but does not validate historical GBM implementation or prove a rate of convergence. Thus the work is best viewed as a strengthened, transparent preprint rather than a claim of a new general convergence theorem.

\section*{Disclosure statement}
No potential conflict of interest was reported by the author.
\section*{Funding}
The research received no specific grant from a public, commercial or not-for-profit funding agency.

\end{document}